\documentclass{jfm}
\usepackage{graphicx}
\usepackage{epstopdf, epsfig}
\usepackage{amsmath}
\DeclareMathOperator{\eig}{eig}
\DeclareMathOperator{\im}{Im}
\DeclareMathOperator{\sign}{sgn}
\usepackage{bigints}
\usepackage{siunitx}
\DeclareSIUnit \px {pixels}
\usepackage{enumitem}
\usepackage{tabularx}
\usepackage{multirow}
\usepackage{hyperref}
\hypersetup{
  colorlinks=true,
  linkcolor=blue,
  urlcolor=blue,
  citecolor=blue,
}

\usepackage{nomencl}
\makenomenclature
\usepackage{etoolbox}
\renewcommand\nomgroup[1]{%
  \item[\it
  \ifstrequal{#1}{A}{Primary symbols}{%
  \ifstrequal{#1}{B}{Vectors}{%
  \ifstrequal{#1}{C}{Secondary symbols}{%
  \ifstrequal{#1}{D}{Dimensionless}{%
  \ifstrequal{#1}{E}{Subscripts}{%
  \ifstrequal{#1}{F}{Statistics}{%
  }}}}}}%
]}
\usepackage{multicol}
\renewcommand{\nompreamble}{\begin{multicols}{2}}
\renewcommand{\nompostamble}{\end{multicols}}
\shorttitle{Role of the Curvature in the Development of Flows Reattached over an Airfoil}
\shortauthor{A. Shirinzad, K. Xu and P. E. Sullivan}

\title{On the Role of the Wall Curvature in the Development of Flows Reattached over an Airfoil through Unsteady Blowing}

\author{Ali Shirinzad\aff{1}
  \corresp{\email{ali.shirinzad@mail.utoronto.ca}},
  Kecheng Xu\aff{2}
 \and Pierre Edward Sullivan\aff{1}}

\affiliation{\aff{1}Department of Mechanical and Industrial Engineering, University of Toronto, Toronto, Ontario, M5S 3G8, Canada
\aff{2}University of Toronto Institute for Aerospace Studies, Toronto, Ontario, M3H 5T6, Canada}

\begin{document}

\maketitle

\begin{abstract}
An array of twelve circular-orifice synthetic jet actuators (SJAs) was used to provide the unsteady forcing required for flow separation control over a National Advisory Committee for Aeronautics (NACA) 0025 airfoil at a chord-based Reynolds number of \SI{100000}{} and an angle of attack of \ang{10}. Two distinct high- and low-forcing frequencies corresponding to the shear layer and wake instabilities were used at an identical blowing strength for flow control. Particle image velocimetry (PIV) was used to measure the velocity fields at the centerline of the airfoil. The results showed the presence of a turbulent shear layer stretching from the edge of the reattached boundary layer to the irrotational flow with an invariant mean spanwise vorticity in the wall-normal direction. It was revealed that the coherent structures for the high-frequency controlled case are advected along the boundary of the rigid-body rotation shear layer and the irrotational flow, whereas for the low-frequency actuation, some structures directly pass through the rigid-body rotation region, disrupting the wall-normal balance of vorticity. Analytical expressions were derived for the variation of the mean spanwise vorticity in the rigid-body rotation region and the curvature-multiplied mean angular momentum in the irrotational flow region based on order-of-magnitude analysis and semi-empirical grounds. The resulting patterns showed an excellent agreement with the measured experimental data.
\end{abstract}


\renewcommand{\nomname}{Nomenclature}
\nomenclature[A]{\(\alpha\)}{Angle of attack}
\nomenclature[A]{\(\beta\)}{Airfoil surface slope}
\nomenclature[A]{\(c\)}{Airfoil chord length}
\nomenclature[A]{\(\nu\)}{Fluid kinematic viscosity}
\nomenclature[A]{\(\rho\)}{Fluid density}
\nomenclature[A]{\(d\)}{Diameter}
\nomenclature[A]{\(f\)}{Instability frequency}
\nomenclature[A]{\(f_e\)}{Excitation frequency}
\nomenclature[A]{\(f_m\)}{Modulated frequency}
\nomenclature[A]{\(V_{pp}\)}{Peak-to-peak voltage}
\nomenclature[A]{\(\phi\)}{Phase angle}
\nomenclature[A]{\(u, v, w\)}{Velocity components}
\nomenclature[A]{\(p\)}{Pressure}
\nomenclature[A]{\(\Gamma\)}{Circulation}
\nomenclature[A]{\(\lambda\)}{Swirling strength}
\nomenclature[A]{\(Q\)}{$Q$-criterion}
\nomenclature[A]{\(X, Y\)}{Global Cartesian coordinates}
\nomenclature[A]{\(x, y, z\)}{Local Cartesian coordinates}
\nomenclature[A]{\(s, n\)}{Curvilinear coordinates}
\nomenclature[A]{\(r_C\)}{Radius of curvature}
\nomenclature[A]{\(\kappa\)}{Curvature}
\nomenclature[A]{\(S\)}{Surface area}
\nomenclature[A]{\(a\)}{Airfoil profile coefficients}
\nomenclature[A]{\(T\)}{Airfoil thickness coefficient}
\nomenclature[A]{\(A\)}{Linear profile slope}
\nomenclature[A]{\(B\)}{Linear profile intercept}
\nomenclature[A]{\(h\)}{Lame coefficient}

\nomenclature[B]{\(\boldsymbol{e}\)}{Unit vector}
\nomenclature[B]{\(\boldsymbol{r}\)}{Position vector}
\nomenclature[B]{\(\boldsymbol{v}\)}{Velocity vector}
\nomenclature[B]{\(\boldsymbol{L}\)}{Angular momentum}
\nomenclature[B]{\(\boldsymbol{\Omega}\)}{Vorticity vector}

\nomenclature[C]{\(\mathcal{C}\)}{Airfoil profile}
\nomenclature[C]{\(\mathcal{O}\)}{Order of magnitude}
\nomenclature[C]{\(\mathcal{K}\)}{Related to curvature gradient}
\nomenclature[C]{\(\mathcal{L}\)}{Related to angular momentum}
\nomenclature[C]{\(\mathcal{R}\)}{Residual function}
\nomenclature[C]{\(\mathcal{F}\)}{Residual fraction}

\nomenclature[D]{\(C_p\)}{Pressure coefficient}
\nomenclature[D]{\(Re\)}{Reynolds number}
\nomenclature[D]{\(St\)}{Strouhal number}
\nomenclature[D]{\(Ma\)}{Mach number}
\nomenclature[D]{\(C_B\)}{Blowing ratio}
\nomenclature[D]{\(F^+\)}{Reduced modulated frequency}
 
\nomenclature[E]{\(0, \cdots, 4\)}{For different coefficients}
\nomenclature[E]{\(\infty\)}{Freestream conditions}
\nomenclature[E]{\(c\)}{Based on chord}
\nomenclature[E]{\(o\)}{Relating to orifice}
\nomenclature[E]{\(x, y, z\)}{Cartesian axes directions}
\nomenclature[E]{\(t, n\)}{Tangent and normal directions}
\nomenclature[E]{\(s\)}{Relating to arc length}
\nomenclature[E]{\(O\)}{Referring to a reference point}
\nomenclature[E]{\(ir\)}{Relating to irrotational flow}
\nomenclature[E]{\(rr\)}{Relating to rigid-body rotation}
\nomenclature[E]{\(re\)}{Relating to residual component}

\nomenclature[F]{$\overline{(.)}$}{Time-averaged value}
\nomenclature[F]{$(.)'$}{Fluctuating component}
\nomenclature[F]{$\langle . \rangle $}{Phase-averaged value}
\nomenclature[F]{$\widetilde{(.)}$}{Coherent component}
\nomenclature[F]{$(.)''$}{Incoherent component}
\nomenclature[F]{\(R^2\)}{Coefficient of determination}
\nomenclature[F]{\(N\)}{Sample size}
\nomenclature[F]{\(Z_C\)}{Confidence coefficient}
\nomenclature[F]{\({\xi _{\overline {u_x} }}, {\xi _{\overline {v_y} }}\)}{Uncertainty in velocities}
\nomenclature[F]{\({\xi _{\overline {u_x' v_y'} }}\)}{Shear stress uncertainty}

\printnomenclature

\section{Introduction}\label{sec:intro}
Flow separation is the breakaway or detachment of fluid from a bounding surface \citep{Gad-el-Hak1991} caused by an adverse pressure gradient \citep{Simpson1989}, a geometrical aberration \citep{Bradshaw1972,Kim1980}, or other means. Flow separation is marked by a thickening of the rotational flow region adjacent to the surface and a significant increase in the wall-normal velocity component \citep{Gad-el-Hak1991,Greenblatt2000}. The performance of flow systems is often controlled by the separation location due to the large energy losses associated with this phenomenon. For example, delaying the separation decreases the pressure drag of a bluff body \citep{Gad-el-Hak1991} or enhances the circulation and hence the lift of an airfoil at
high angles of attack \citep{Amitay2002a}. For airfoils in particular, it is important to develop control strategies to mitigate the separation, as this geometry forms the cross-section of lifting bodies used in several applications ranging from aircrafts to power-generating turbines.

\subsection{Flow separation over an airfoil}\label{subsec:intro_separation}
Flow separation over an airfoil is seen in many engineering applications, including low-pressure turbine (LPT) blades operating at high altitudes, micro-air vehicles (MAVs), and unmanned aerial vehicles (UAVs) \citep{Hodson2005,Shkarayev2007,Yang2008,Savaliya2010}. A schematic of flow over an airfoil is shown in figure~\hyperref[fig:nom_a]{\ref{fig:nom_a}(a)}, where $\alpha$ is the angle of attack, $c$ is the chord length, and $u_\infty$ is the freestream velocity. In the context of flow over an airfoil, the chord-based Reynolds number is defined as $Re_c=u_\infty c/\nu$ where $\nu$ is the fluid kinematic viscosity. A comprehensive summary of the flow phenomenon for Reynolds numbers ranging from \SI{1000}{} to \SI{200000}{} is reported by \citet{Carmichael1981}. Generally, the Reynolds number is considered low when $Re_c<10^6$, particularly when $Re_c<\SI{50000}{}$ \citep{Carmichael1981,Lissaman1983,Fitzgerald1990}. For airfoils operating at low Reynolds numbers, the momentum of the laminar boundary layer is not sufficient to withstand the adverse pressure gradient on the suction surface, and consequently, the flow often separates near the leading edge of the airfoil to form a separated shear layer \citep{Winslow2018}.

Experimental and numerical investigations have confirmed two distinct flow regimes for low-Reynolds-number flow over airfoils. At sufficiently low  Reynolds numbers, the separated shear layer fails to reattach to the airfoil surface, undergoing laminar-to-turbulent transition due to amplification of flow instabilities before forming a turbulent wake \citep{Carmichael1981,Hoarau2003}. This outcome, which is often referred to as stall, causes a significant decrease in lift-to-drag ratio. In general, the angle of attack has a similar but inverse effect on the state of the flow \citep{Yang2008,Winslow2018}. Hence, in this Reynolds number range, flow separation tends to occur at a lower angle of attack, which limits the effective operation range of the airfoil \citep{Yarusevych2009,Boutilier2012a}. At sufficiently large Reynolds numbers, the turbulent separated shear layer may reattach to the airfoil surface, resulting in a laminar separation bubble (LSB) \citep{Winslow2018}. The transition to turbulence for an LSB follows the same process as in the case of a separated shear layer \citep{Dovgal1994,Boutilier2012b}. At a constant angle of attack, the transition between stall and a laminar separation bubble has been shown to be an unsteady phenomenon that occurs over a finite range of Reynolds numbers \citep{Carmichael1981}. Laminar separation bubbles are also detrimental to aerodynamic performance, even when the bubble is relatively short and spans a small percentage of the airfoil chord \citep{Gaster1967,Fitzgerald1990}. The unsteady aerodynamic loads inherent in both stalled flow and laminar separation bubbles cause structural vibrations, noise, and fatigue failure \citep{Yarusevych2009}. Given the undesirable effects associated with flow separation over airfoils, boundary layer development and separation control have been the focus of many of the studies over the past few decades \citep{Tani1964,Lissaman1983,Lin1996}.

\subsection{Synthetic jet actuators in a quiescent ambient}\label{subsec:intro_quiescent}
\begin{figure}
  \centerline{\includegraphics[width=\linewidth, trim={2, 2, 2, 2}, clip]{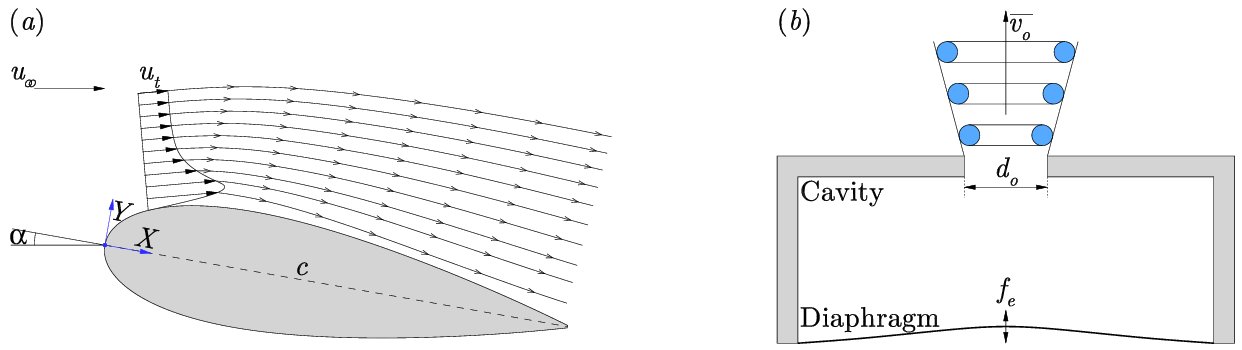}}
  \caption{Schematic representations and the pertinent parameters for (\textit{a}) flow over an airfoil and (\textit{b}) a synthetic jet actuator (SJA).}
\label{fig:nom_a}
\end{figure}

Flow control devices are generally classified into active and passive devices. Passive flow control devices, such as vortex generators on the wings of most passenger aircrafts, delay flow separation without requiring energy \citep{Debien2016}. In contrast, active flow control devices, such as fluidic jets, require input power to add momentum to the flow. However, they can adapt to off-design flow conditions, and they do not introduce a drag penalty common with passive control devices permanently mounted on the airfoil. A synthetic jet actuator (SJA) is an active control device consisting of a vibrating diaphragm that alters the air volume inside a cavity to produce a synthetic jet through an orifice. The periodic oscillation of the diaphragm leads to a cycle of ingestion and expulsion of the working fluid, forming a train of vortex structures that propagate away from the orifice \citep{Glezer2002}. A schematic representation of an SJA is presented in figure~\hyperref[fig:nom_a]{\ref{fig:nom_a}(b)}, where $f_e$ is the diaphragm excitation frequency, $d_o$ is the orifice diameter, and $\overline{v_o}$ is the the time-averaged blowing velocity at the orifice centerline. As the synthetic jet leaves the cavity, low-momentum fluid enters through the perimeter of the orifice to replace fluid ejected by the diaphragm. Hence, these devices are non-zero-net momentum-flux but zero-net mass-flux (ZNMF). ZNMF SJAs significantly reduce the complexity and weight of the flow control systems since they require no bleed air supply or ducting. Periodic forcing provided by SJAs has been observed to offer greater entrainment of fluid in the near field compared to continuous blowing methods \citep{James1996,Smith2001,Cater2002}. Furthermore, the cost of unsteady blowing is less than that of steady methods, and in some instances, the difference is observed to be an order of magnitude \citep{Wygnanski1997}, making SJAs a viable solution for active flow control applications.

SJAs are usually characterized in a quiescent ambient to unravel the relationship between the blowing velocity $\overline{v_o}$, actuation frequency $f_e$, and the voltage applied to the diaphragm $V_{pp}$. The effects of a wide range of parameters, such as orifice shape, blowing velocity, and actuation frequency, on the flow characteristics of synthetic jets have been studied extensively. A parametric study performed by \citet{Jabbal2006} revealed that the synthetic jet flow in a quiescent ambient is characterized by the Strouhal number $St=f_e d_o/\overline{v_o}$ and the jet Reynolds number $Re=\overline{v_o} d_o/\nu$. Still, the evolution of the flow structures near and far away from an SJA is highly dependent on the orifice shape \citep{Lindstrom2019}. \citet{Amitay2006}, considered slit-like orifices with aspect ratios from 50 to 100 to have a finite span. \citet{Sahni2011} also classified SJAs with a slit-like orifice based on their aspect ratio into high-aspect-ratio slits having aspect ratios of 75 or above, low-aspect-ratio slits with aspect ratios below 5, and finite-span or rectangular slits with end effects having aspect ratios between 5 and 75. Both experimental and numerical studies have identified the flow structures associated with a slit-like orifice as periodically-formed two-dimensional discrete vortex pairs near the slit that undergo spanwise instability before transitioning to turbulence, ultimately losing coherence at some location downstream of the orifice \citep{Smith1998,Rizzetta1999,Yao2004}. The experimental study led by \citet{Amitay2006} for a finite-span synthetic jet showed that while the issued synthetic jet is two-dimensional near the orifice, secondary counter-rotating structures are formed farther downstream, which were attributed to the orifice edge effects. Similar observations were made for three-dimensional flow structures and instabilities for synthetic jets issued through a circular orifice \citep{Crook2001,Cater2002}.

\subsection{Synthetic jet actuators in a crossflow}\label{subsec:intro_crossflow}
For flow control applications, the effective performance of SJAs significantly depends on the interactions of the synthetic jet with a crossflow. Both high and low-aspect-ratio SJAs have been investigated in the presence of a crossflow, though low-aspect-ratio SJAs are often grouped into an array to cover a longer span. Based on the parametric studies and dye visualizations performed by \citet{Jabbal2008}, \citet{Zhang2010}, and \citet{Zhong2013}, in the presence of a crossflow the vortical structures issued from the SJA orifice are significantly affected by the jet-to-freestream velocity ratio $C_B=\overline{v_o}/u_\infty$, also called the blowing ratio. These studies identified a minimum blowing ratio below which the synthetic jets cannot penetrate the crossflow. It was observed that above this threshold, the primary vortical structures produced by the interaction between a round synthetic jet and a boundary layer transition from hairpin-like vortices located close to the wall to tilted vortex rings that penetrate the edge of the boundary layer. According to \citet{Jabbal2008}, the hairpin-like structures are a result of the upstream branch of the vortex ring produced by the circular SJA being pulled into the cavity during the suction cycle, whereas the tilted vortex rings emerge when the synthetic jets are able to escape from the ingestion of the suction cycle at high enough blowing ratios. Surface flow visualizations of \citet{Zhang2010} showed that both the hairpin-like vortex and the tilted vortex ring are connected to the orifice via two counter-rotating legs, which induce streamwise vortical structures that create a downwash and entrain high-speed fluid from the freestream to the near-wall region. The hairpin-like vortices and tilted vortex rings for low-Reynolds number flows have also been seen in computational fluid dynamics (CFD) simulations \citep{Zhou2009,Palumbo2022}.

Periodic excitation interferes with the growth of flow instabilities within a crossflow, highlighting the importance of the Strouhal number defined as $St_c=f c/u_\infty$, where $f$ is the frequency of flow instabilities \citep{Greenblatt2000}. Collectively, the previous works indicate that active flow control schemes that target the dominant flow instabilities can significantly delay flow separation \citep{Chang1992,Seifert1996,Amitay2002b,Glezer2005,Deem2020}. The post-stall separated flow over an airfoil is dominated by two major flow instabilities, namely the global instability that causes large-scale vortex shedding in the wake and the local instability of the separated shear layer \citep{Wu1998,Yarusevych2009}. The presence of a laminar separation bubble introduces a third major instability, that is the bubble flapping or shedding frequency \citep{Raju2008,Marxen2011,Deem2020}. For the post-stall flow over an airfoil, the dimensionless frequencies corresponding to the shear layer and wake instabilities are an order of $St_c \sim \mathcal{O}(10)$ \citep{Brendel1988,Boutilier2012b} and $St_c \sim \mathcal{O}(1)$ \citep{Yarusevych2006,Buchmann2013}, respectively. The difference in the order of magnitude of these reduced frequencies follows from conventional scaling arguments, where the characteristic length scale in the shear layer near separation is an order of magnitude smaller than that of the wake \citep{Tian2006}. For SJAs mounted on an airfoil, periodic forcing may be applied using the conventional time-harmonic actuation or pulse-modulated actuation at a dimensionless modulated frequency $F_c^+=f_m c/u_\infty$, which may be useful in situations when either the location or the blowing strength of the SJAs are sub-optimal \citep{Amitay2002a}. In the study of \citet{Amitay2002a}, the pulse modulation of SJAs led to a substantial lift recovery when compared to the time-harmonic actuation at the same blowing strength. Experiments indicate that forcing near the wake frequency leads to unsteady reattachment and aerodynamic forces accompanied by large vortex formation and advection, whereas actuation near the shear layer frequency results in a more steady flow reattachment altering the local pressure gradient to suppress flow separation \citep{Amitay2002b,Glezer2005}. For example, \citet{Amitay2002a} observed that the transients following the initiation or termination of the pulse-modulated control are very similar for $F_c^+ \sim \mathcal{O}(10)$ and $F_c^+ \sim \mathcal{O}(1)$ cases. Following the initial transition, the shedding of organized vortical structures eventually subsided for $F_c^+ \sim \mathcal{O}(10)$ case, whereas $F_c^+ \sim \mathcal{O}(1)$ case was accompanied by the coherent shedding of a train of large vortices. These observations were confirmed in a more recent study by \citet{Xu2023} who used an array of circular microblowers to control separation over a NACA0025 airfoil.

Several researchers have considered the effects of actuation parameters on the three-dimensionality of the resulting controlled flow. \citet{Sahni2011} investigated the three-dimensional interactions between an array of finite-span slit-like SJAs and the crossflow over a NACA 4421 airfoil for a range of blowing ratios. At low blowing ratios, the development of spatial non-uniformities due to the finite span of the slit led to the formation of small and organized secondary structures. Increasing the blowing ratio resulted in increased penetration of the jet into the crossflow, increased spanwise wavelength of the secondary structures, and reduced spanwise extent of the interaction domain. The formation of the secondary vortex structures due to the interaction of finite-span synthetic Jets and a crossflow over an airfoil has also been reported by \citet{Vasile2013}. Using tuft and oil visualization methods, \citet{Feero2017b} showed that the spanwise extent of the reattached flow narrows towards the trailing edge of the airfoil, describing this phenomenon as flow contraction towards the airfoil centerline. By varying both the excitation frequency and blowing ratio, they concluded that the spanwise extent of the controlled flow region increases as the blowing ratio increases. These three-dimensional visualizations indicate that the time-averaged controlled flow within a finite extent of the airfoil centerline may be considered quasi-two-dimensional.

\subsection{Effects of curvature}\label{subsec:intro_curvature}
From the discussion in \S\ref{subsec:intro_separation}, \S\ref{subsec:intro_quiescent}, and \S\ref{subsec:intro_crossflow}, it is clear that a considerable number of experimental and numerical studies have already been conducted to study the effect of various SJA parameters, such as orifice shape, blowing ratio, and frequency, on reattached flow over an airfoil. In comparison, the available literature on the effects of curvature is very limited. It is well known that curvature may initiate the generation of streamwise vortices through a centrifugal instability of the flow \citep{Taylor1923}. For convex surfaces, the streamlines strongly adhere to the wall, a phenomenon commonly referred to as the Coanda effect. The crossflow over a Coanda cylinder, which is a two-dimensional constant curvature circular cylinder, is a problem of significant practical and theoretical interest \citep{Greenblatt2000}. \citet{Neuendorf1999} considered the evolution of a two-dimensional wall jet over a Coanda cylinder and observed that the wall jet spreads outward to a greater extent compared to a flat surface, which was attributed to centrifugal instability. \citet{Greenblatt2000} performed an extensive literature review on the effects of the reduced frequency on separated flow over a circular cylinder, reporting that the curvature does not have a significant effect on the optimum reduced frequencies. More recently, \citet{Shirinzad2023a} conducted experiments for turbulent free-surface flows over a circular cylindrical segment with constant curvature and revealed the importance of angular momentum and vorticity, characteristics which were highlighted in the studies on rotating flows \citep{Rayleigh1917,Taylor1923}. In their work, the oncoming flow that was separated at the leading edge due to a geometrical aberration was eventually reattached to the wall to form a recirculation bubble. The examined flow exhibited mixed characteristics of two important canonical flows, namely the flow inside and outside of a revolving circular cylinder. By analyzing the vorticity, a rigid-body rotation shear layer was identified that stretched between the edge of a thin turbulent region beneath the concave free surface and the irrotational flow region above the wall. Analysis of the angular momentum, on the other hand, showed that the angular momentum becomes invariant in the wall-normal direction within the irrotational flow region over the reattached boundary layer, even in the absence of axial symmetry. These empirical observations allowed the governing equations to be simplified to characterize the angular momentum within the irrotational flow region.

The above discussion clearly shows that the effects of wall curvature are commonly studied using a constant curvature body. For an airfoil equipped with an array of SJAs, the initiation of actuation is known to cause a Coanda-like attachment of the separated shear layer \citep{Amitay2002a}. Still, airfoils have varying curvature bodies, and to the best of the authors' knowledge, there are no studies on how the reattached flow properties change with respect to the wall curvature. Hence, the present work is motivated by the need for an in-depth understanding of the effects of varying curvature on flow properties over an airfoil. The flow is primarily studied at the centerline of the airfoil, where it may be considered quasi-two-dimensional as explained in \S\ref{subsec:intro_crossflow}. It should be noted that the development of the reattached flow at the centerline is of particular interest as the centerline flow represents the best achievable control given a set of actuation parameters. The objectives of this paper are therefore as follows:
\begin{enumerate}[label = (\roman*), align=left, itemindent=0pt, leftmargin=*, widest=iii]
\item Investigate the rotational characteristics, namely vorticity and angular momentum, for turbulent flows over a broader range of geometries having varying curvature.
\item Investigate the effects of the forcing frequency of an array of circular SJAs on curvature-dependent flow characteristics, such as vorticity and Reynolds stresses near the wall.
\end{enumerate}
The remainder of this paper is organized as follows. The experimental setup, data processing, data post-processing, and the analytical approach are described in \S\ref{sec:procedure}. The main results and the pertaining discussion are presented in \S\ref{sec:results}, while the major findings and conclusions are summarized in \S\ref{sec:summary}. Supplementary analytical tools and discussion are included in appendices~\ref{app:uncertainty}-\ref{app:governing_equations}.

\section{Methodology}\label{sec:procedure}
\subsection{Wind tunnel facility}
\begin{figure}
  \centerline{\includegraphics[width=\linewidth, trim={2, 2, 2, 2}, clip]{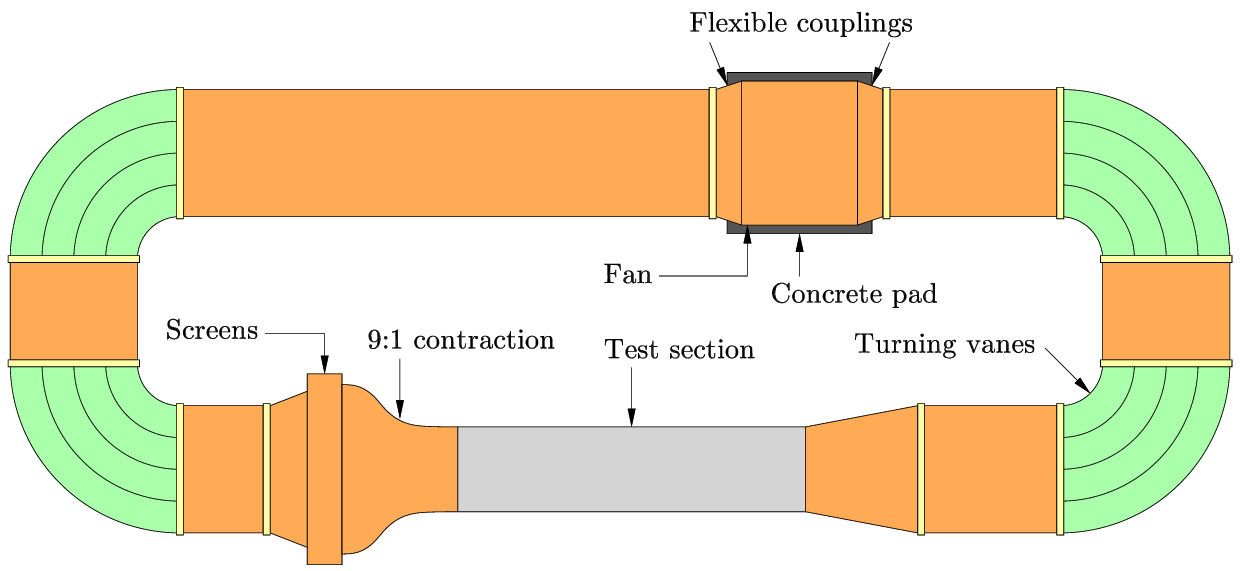}}
  \caption{A schematic showing various components of the wind tunnel facility.}
\label{fig:nom_b}
\end{figure}

The experiments were conducted in a closed return low-speed wind tunnel located in the Turbulence Research Laboratory (TRL) at the Department of Mechanical and Industrial Engineering, University of Toronto. A schematic showing the main components of the wind tunnel facility is shown in figure~\ref{fig:nom_b}. The test section of the wind tunnel is \SI{5000}{\milli\meter} long with an octagonal cross-section \SI{1220}{\milli\meter} high and \SI{910}{\milli\meter} wide. The corners of the octagonal cross-section have a constant angle but decrease in width along the test section length to increase the cross-sectional area and compensate for boundary layer growth. The ceiling and one of the side walls of the test section are fabricated from clear acrylic plates to facilitate optical access. The flow in the tunnel is driven by a six-bladed axial fan, powered by a REEVES\textsuperscript{\textregistered} MotoDrive\textsuperscript{\textregistered} 500 series motor located outside the wind tunnel on an isolating concrete pad. The fan housing is connected to the wind tunnel by flexible couplings to minimize the transfer of vibrations from the fan to the tunnel structure. The freestream velocity in the test section is adjustable from \SI{2.5}{\meter/\second} to \SI{18.0}{\meter/\second}, monitored using a pitot-static tube installed at the test section inlet with an uncertainty estimated to be less than $\pm\SI{1}{\percent}$ \citep{Xu2023}. The flow entering the test section passes through a conditioning unit consisting of seven screens and a 9:1 converging section to minimize the turbulence and homogenize the flow. The freestream is mostly uniform with a turbulence intensity generally less than \SI{0.1}{\percent}. The flow exiting the test section is redirected by four \ang{90} corners of the wind tunnel through a turning vane system for flow recirculation.

\subsection{Airfoil model and instrumentation}
The airfoil model used in the present study had a NACA 0025 profile with an open trailing edge, the same as the one used by \citet{Xu2023}. The airfoil model was machined from aluminum, having a chord length of $c=\SI{300}{\milli\meter}$ and a spanwise extent of \SI{885}{\milli\meter}. Approximately 1/3 of the middle section of the model is hollow to permit the installation of SJAs and sensors. The model was fitted with circular acrylic end plates to suppress edge effects and the influence of the tunnel sidewall boundary layer. It has been verified experimentally that end effects do not influence flow development over at least \SI{50}{\percent} of the airfoil span within the domain of interest \citep{Yarusevych2009}. The model with the attached end plates was installed \SI{400}{\milli\meter} downstream of the entrance to the test section, spanning the entire width of the cross-section. A rotation lock and bearing housings attached to the end plates were used to adjust the pitch angle of the airfoil. A digital protractor was employed to set the angle of attack to the desired values with an uncertainty of $\pm\ang{0.1}$ \citep{Yarusevych2009,Xu2023}.

An array of 24 Murata MZB1001T02 microblowers, distributed equally in two rows and symmetrically around the airfoil centerline, was mounted near the leading edge inside the hollow section of the airfoil using a rectangular housing \SI{317}{\milli\meter} wide and \SI{58}{\milli\meter} long. The upstream row of the array was located at $X_j/c = 0.10$, just upstream of the separation point $X/c=0.12$ reported by \citet{Xu2023}. Generally, separation control is most effective when the excitation is applied in the vicinity of the separation point, yet not to the stable flow near the leading
edge \citep{Greenblatt2000}. The SJA array was wired such that each row was powered independently, allowing the use of only the upstream row needed in the present study. Each SJA had a circular orifice of diameter $d_o=\SI{0.8}{\milli\meter}$ the center of which was \SI{25}{\milli\meter} apart from the center of the nearest microblowers. The microblowers operational range was between \SI{5}{\volt} to \SI{30}{\volt}, capable of providing \SI{24.0}{\kilo\hertz} to \SI{27.0}{\kilo\hertz} of excitation frequency. The supplied signal to the microblowers was a square wave with \SI{50}{\percent} duty cycle and adjustable burst-modulated frequency, created using a Rigol DG1022Z function generator with a square carrier wave at a frequency of $f_e=\SI{25.5}{\kilo\hertz}$, which was amplified by a YAMAHA HTR5470 amplifier to a peak-to-peak voltage within the microblower operational range.

\subsection{Velocity and pressure measurements}
The experiments involved three test conditions, namely the uncontrolled flow, high-frequency forcing $F_c^+ \sim \mathcal{O}(10)$, and low-frequency forcing $F_c^+ \sim \mathcal{O}(1)$, at a chord-based Reynolds number of $Re_c=10^5$ and an angle of attack of $\alpha=\ang{10}$. The chord-based Reynolds number was then set to the desired value by keeping the freestream velocity relatively constant at $u_\infty=\SI{5.2}{\meter/\second}$. The resulting Mach number at this freestream velocity was low enough for the flow to be considered incompressible, that is $Ma \ll 0.3$. At the examined Reynolds number, post-stall flow conditions were established for the uncontrolled flow. For both controlled cases, the peak-to-peak voltage of the input signal to the SJA array was set to $V_{pp}=\SI{20}{\volt}$, corresponding to $\overline{v_o}=\SI{25.0}{\meter/\second}$ in a quiescent ambient and a blowing ratio of $C_B=4.81$ \citep{Xu2023}. Meanwhile, the burst-modulated frequency was adjusted to $f_m=\SI{200}{\hertz}$ and $\SI{20}{\hertz}$, resulting in a reduced frequency of $F_c^+=11.54$ and $1.15$, respectively.

The suction side of the airfoil model was equipped with thirty-three pressure taps \SI{0.8}{\milli\meter} in diameter, located at the airfoil model centerline connected to a Scanivalve pressure scanner through pneumatic tubing. The pressures were measured using an MKS Baratron\textsuperscript{\textregistered} 226A pressure transducer with a range of $\pm\SI{26.66}{\pascal}$ in conjunction with the Scanivalve pressure scanner. The maximum uncertainty associated with the surface pressure measurements has been shown to be $\pm\SI{2}{\percent}$ \citep{Yarusevych2009}. At each port, a total of \SI{30000} samples were collected at a frequency of \SI{1000}{\hertz} to resolve the time-averaged surface pressure distribution along the airfoil. The freestream static pressure $p_\infty$ was also measured from the static side of the pitot tube used to measure the freestream velocity at the test section inlet.

\begin{figure}
  \centerline{\includegraphics[width=\linewidth, trim={2, 2, 2, 2}, clip]{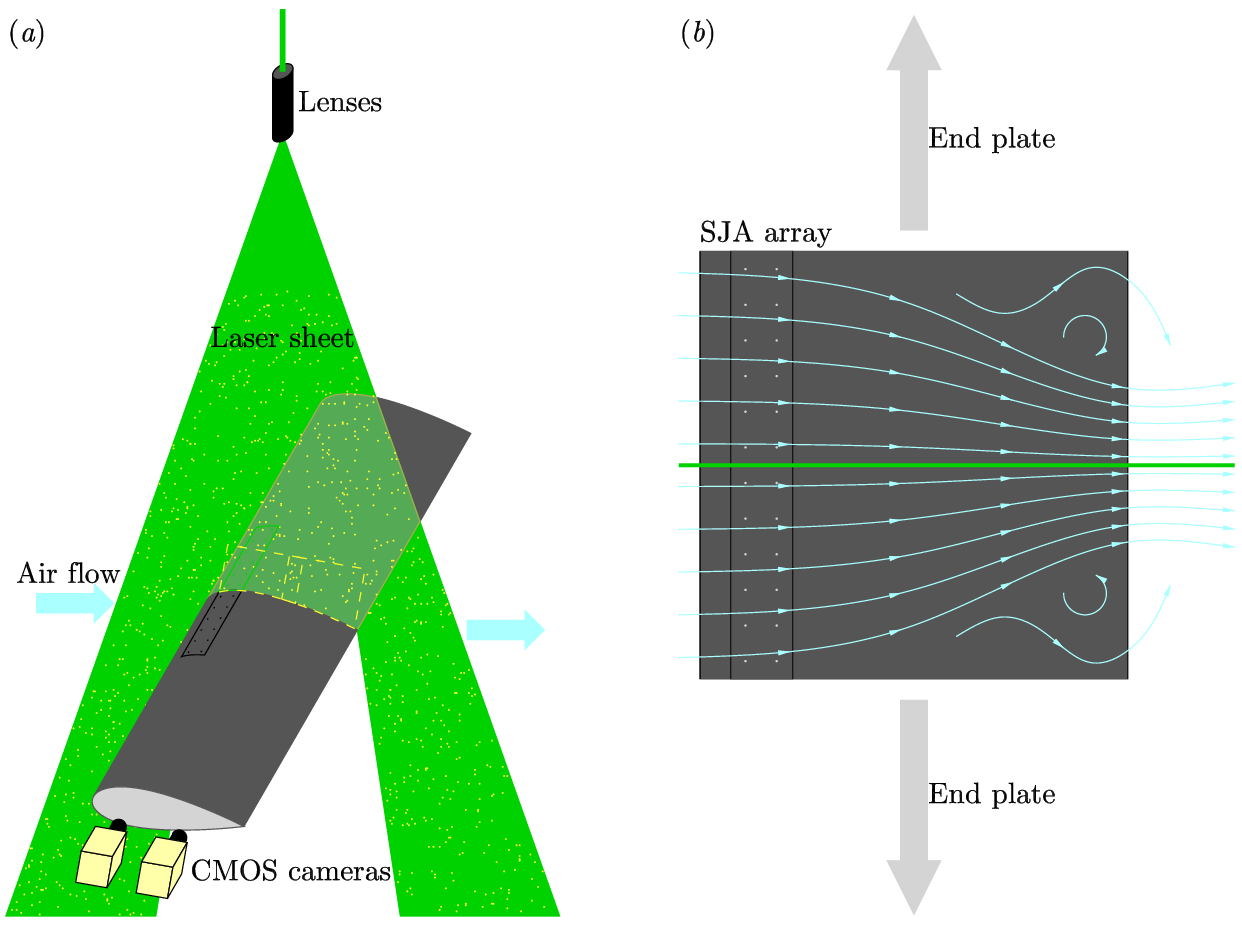}}
  \caption{A schematic showing (\textit{a}) pitched airfoil model, rotated cameras, and the resulting fields of view and (\textit{b}) top view of the section of the airfoil equipped with the microblowers and the location of the aligned laser light sheet.}
\label{fig:nom_c}
\end{figure}

A two-dimensional two-component (2D2C) particle image velocimetry (PIV) system was employed to measure the instantaneous velocity fields at the spanwise centerline of the airfoil model. The flow was seeded using a SAFEX\textsuperscript{\textregistered} 2010F fog generator with SAFEX\textsuperscript{\textregistered}-Inside-Nebelfluid, which is a mixture of diethylene glycol and water. A circular beam was generated by a Litron Bernoulli neodymium-doped yttrium aluminum garnet (Nd:YAG) laser capable of emitting green light up to a maximum pulse energy of \SI{200}{\milli\joule/pulse} at a wavelength of \SI{532}{\nano\meter}. The laser beam was redirected over the test section ceiling where it was spread into a light sheet by concave and convex THORLABS cylindrical lenses with a focal length of \SI{-13.7}{\milli\meter} and \SI{1000}{\milli\meter} to illuminate the seeding particles. The laser light sheet had a thickness of approximately \SI{1}{\milli\meter}, and was carefully aligned at the airfoil centerline to minimize perspective errors arising from the spanwise contraction of the controlled flows. Due to the presence of the pressure taps, the airfoil model was sprayed black to minimize reflection, and no further anti-reflection material could be applied. Two 12-bit complementary metal oxide semiconductor (CMOS) JAI SP5000M-USB cameras fitted with Azure 5022ML12M \SI{50}{\milli\meter} lenses were positioned side-by-side aligned with the airfoil chord to capture the light scattered by the illuminated seeding particles within the fields of view. Both cameras had a resolution of $\SI{2560}{\px} \times \SI{2048}{\px}$ and a pixel density of $\SI{17.1}{\px/\milli\meter}$ after calibration. The resulting field of view for both cameras were \SI{149.7}{\milli\meter} long and \SI{119.8}{\milli\meter} high, overlapping by \SI{30.0}{\milli\meter} in the chordwise direction. A schematic showing the airfoil model, SJA array, and the PIV arrangements is provided in figure~\ref{fig:nom_c}. The image acquisition was timed by an NI PCI-6232e data acquisition card at a sampling frequency of \SI{10}{\hertz} to obtain statistically independent samples. Following \citet{Scharnowski2019} and \citet{Xu2023}, the time delay between the two frames in an image pair was set to $\SI{120}{\micro\second}$ to obtain an appropriate in-plane particle displacement. For the controlled cases, the image recording was synchronized to the start of the actuation cycle to allow for phase-locked data acquisition. For each phase-locked case, starting at $\phi=\ang{0}$ and ending with $\phi=\ang{315}$ with an increment of $\phi=\ang{45}$, 1000 image pairs were collected, resulting in a total of 8000 image pairs. For the post-stall flow, 1000 image pairs were acquired without a particular reference event in the flow.

The images from the two cameras were stitched using a linear combination of the intensities in the overlapping region, resulting in images of size $\SI{4603}{\px} \times \SI{2052}{\px}$. A mask was created to cover the airfoil and a thin region of approximately $\SI{15}{\px}$ above the airfoil surface, which was contaminated by the reflection. All images were transferred to the Niagara supercomputing cluster at the SciNet High-Performance Computing consortium, where they were processed with the open-source software OpenPIV-Python-CPU, utilizing a window deformation iterative multigrid (WIDIM) algorithm \citep{Shirinzad2023b}. The PIV process comprised of an initial iteration at a window size of $\SI{64}{\px} \times \SI{64}{\px}$, two iterations at $\SI{32}{\px} \times \SI{32}{\px}$, and two iterations at a final window size of $\SI{16}{\px} \times \SI{16}{\px}$. The resulting vector fields were $\SI{269}{\milli\meter} \times \SI{120}{\milli\meter}$ large with a spacing of \SI{0.47}{\milli\meter}. Matlab\textsuperscript{\textregistered} and Python were used for data post-processing and calculating the phase-averaged velocity fields, mean values, and higher-order moments of statistics. A statistical convergence test and measurement uncertainty for several statistics are provided in appendix~\ref{app:uncertainty}. All data analysis and visualization were accomplished using the commercial software Origin\textsuperscript{\textregistered}.

\section{Results and discussion}\label{sec:results}
The behavior of certain flow characteristics only becomes apparent when observed in a proper coordinate system. In this study, three distinct coordinate systems were adopted, namely the global Cartesian coordinates $(X,Y)$ shown in figure~\hyperref[fig:nom_a]{\ref{fig:nom_a}(a)}, the local Cartesian coordinates $(x,y,z)$ employed during PIV processing, and the curvilinear coordinates $(n,s,z)$ that uses the airfoil profile as its reference curve to locate any point in the measurement plane using the arc length $s$ and the wall-normal distance $n$. A set of analytical tools needed for studying the flow characteristics in curvilinear coordinates, the relation between the three coordinate systems, and a set of expressions used for obtaining the mean velocities and the Reynolds shear stress in the curvilinear coordinates are provided in appendices~\ref{app:curvilinear_coordinates} and \ref{app:coordinate_transformation}. The three velocity components along the axes of each coordinate system are distinguished using the corresponding axis as a subscript, with the exception of the tangential velocity in the curvilinear coordinates, which is commonly denoted by the $t$ subscript instead of $s$. In this section, contour plots as well as both $n$- and $s$-constant one-dimensional profiles are presented to visualize the flow structures and provide detailed measurements of the flow characteristics. The $s$-constant profiles evaluated at eight successive locations, starting at $s/c = 0.1$ and ending at $s/c = 0.8$ with an increment of 0.1, were staggered next to one another while limiting the range of the normal axis to $0 \le n/c \le 0.28$ to better visualize the evolution of various flow characteristics. The $n$-constant profiles are plotted at several wall-normal locations with a color palette presented on the right-hand side of the plots to facilitate visualization. All plots presented henceforth are normalized by the freestream velocity $u_\infty$ and the airfoil chord $c$.

\subsection{Overview of the reattached flow development}\label{subsec:res_overview}
\begin{figure}
  \centerline{\includegraphics[width=\linewidth, trim={2, 2, 2, 2}, clip]{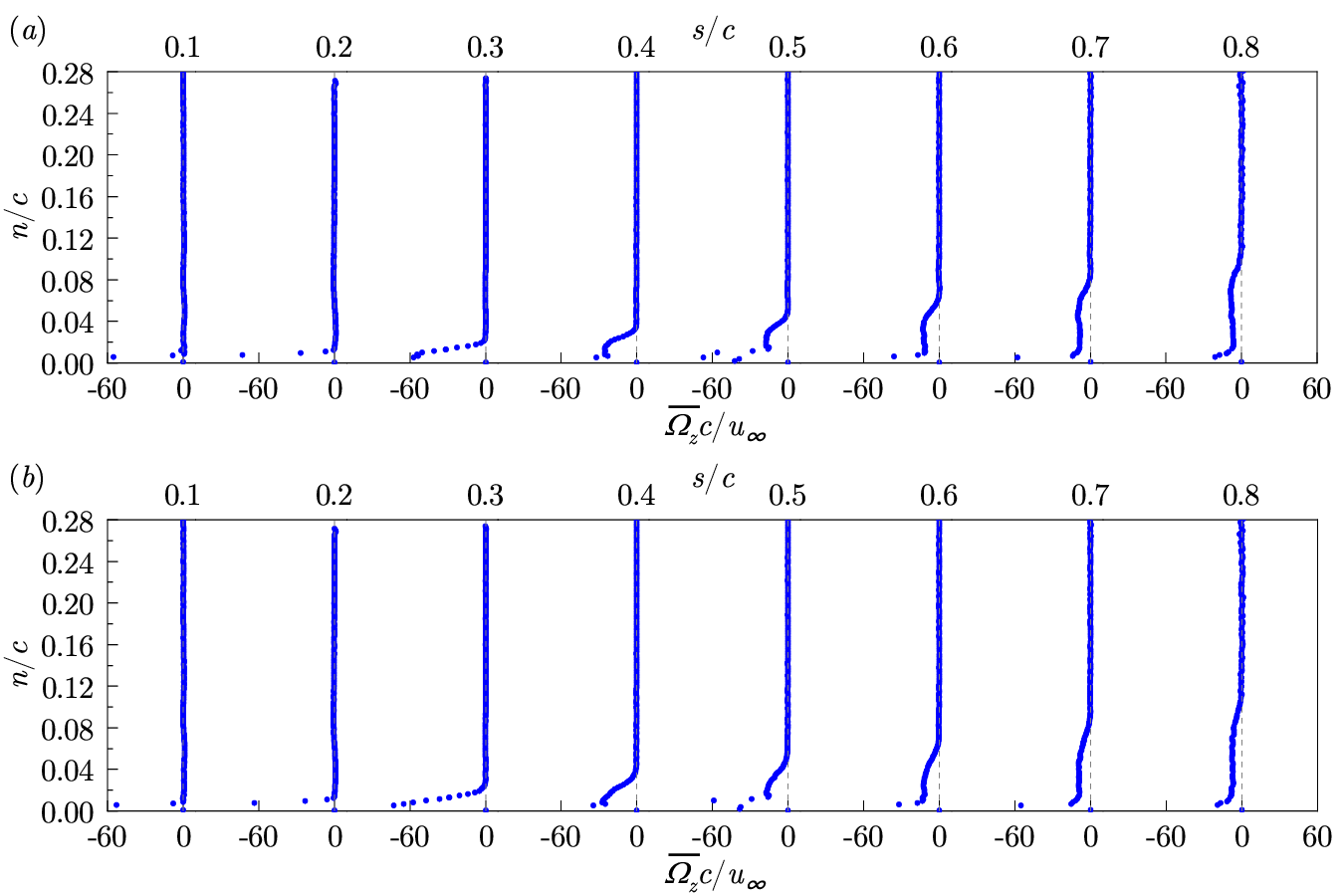}}
  \caption{Tangential evolution of the wall-normal profiles of the mean spanwise vorticity for (\textit{a}) $F_c^+=11.53$ case and (\textit{b}) $F_c^+=1.15$ case.}
\label{fig:Oz_mean}
\end{figure}

\begin{figure}
  \centerline{\includegraphics[width=\linewidth, trim={2, 2, 2, 2}, clip]{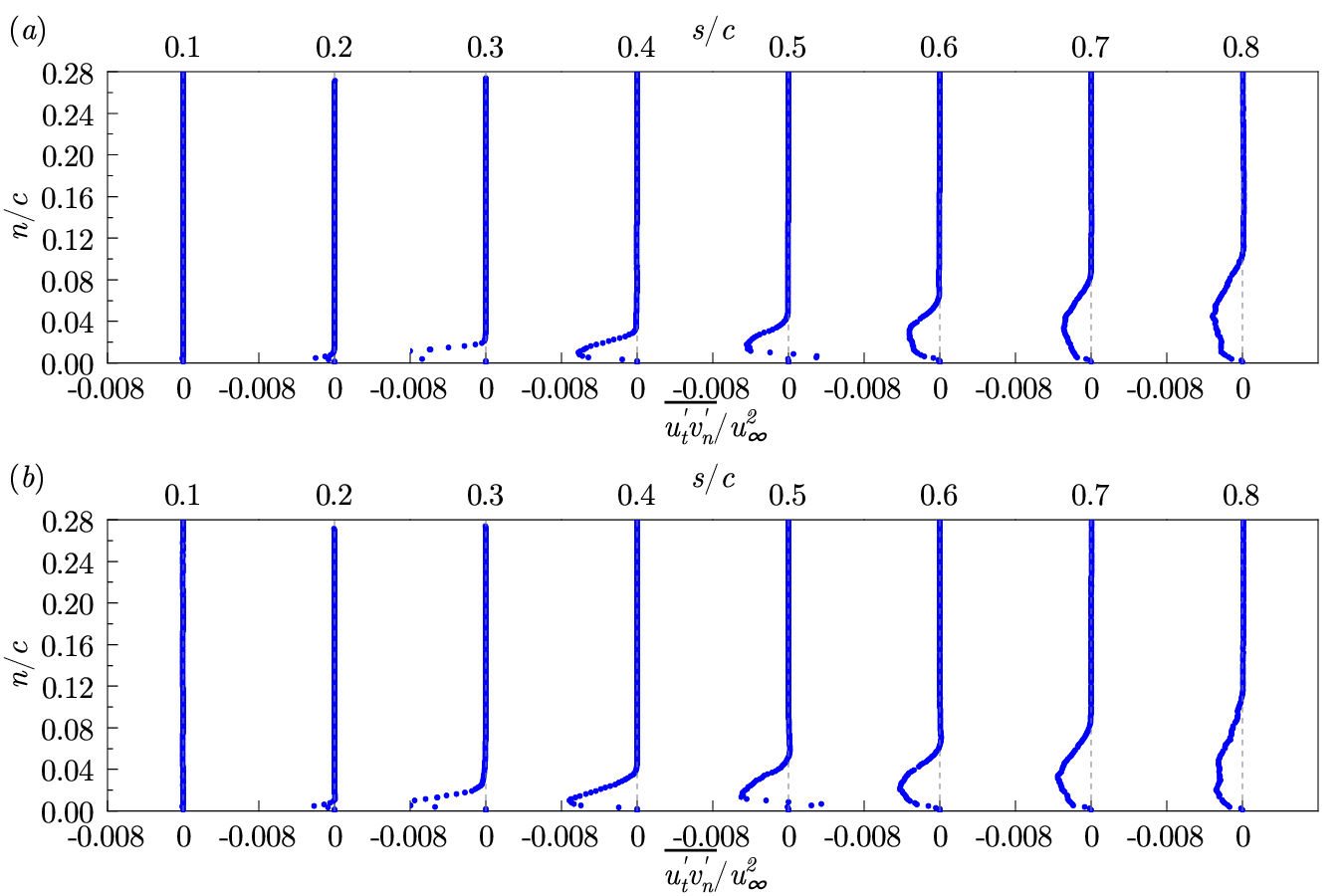}}
  \caption{Tangential evolution of the wall-normal profiles of the mean spanwise vorticity for (\textit{a}) $F_c^+=11.53$ case and (\textit{b}) $F_c^+=1.15$ case.}
\label{fig:uv_mean}
\end{figure}

The study of \citet{Xu2023} revealed that there is no significant difference between the mean velocity fields of the controlled cases. Therefore, we first evaluate the rotational properties of the mean reattached flows in the curvilinear frame. The mean spanwise vorticity was directly calculated from the Cartesian velocities using equation~\eqref{eq:vorticity_cartesian} below to minimize computational errors:
\begin{equation}
\overline {{\Omega _z}}  = \frac{{\partial \overline {{v_y}} }}{{\partial x}} - \frac{{\partial \overline {{u_x}} }}{{\partial y}}
\label{eq:vorticity_cartesian}
\end{equation}
The evolution of the mean spanwise vorticity and the Reynolds shear stress in the wall-tangent direction is presented in figures~\ref{fig:Oz_mean} and \ref{fig:uv_mean}, respectively. An essential feature of turbulent flows is that they are rotational, manifesting as elevated regions of the mean spanwise vorticity and Reynolds shear stress within the reattached boundary layer along the airfoil surface. Two additional regions are also identified in figures~\ref{fig:Oz_mean} and \ref{fig:uv_mean}:

\setlist{noitemsep}
\begin{enumerate}[label = (\roman*), align=left, itemindent=*, leftmargin=0pt, widest=iii]
\item A mostly irrotational region far from the wall, where the normalized vorticity and the normalized Reynolds shear stress are at most an order of $\mathcal{O}(10^{-1})$ and $\mathcal{O}(10^{-5})$, respectively. In this study, this region is referred to as the irrotational flow region.
\item A region close to the wall, extending from the edge of the reattached boundary layer to the irrotational flow region, where the vorticity is almost invariant in the wall-normal direction and the normalized Reynolds shear stress is at most an order of $\mathcal{O}(10^{-2})$. Henceforth, this region is referred to as the rigid-body rotation shear layer.
\end{enumerate}

The existence of the rigid-body rotation shear layer, which was identified by evaluating the spanwise vorticity in the curvilinear frame, is not reported in the previous works on reattached flows over an airfoil. The high turbulence levels of this shear layer and its presence near the wall make it easy to be mistaken with the boundary layer. Similarly, the irrotational flow region should not be mistaken for the freestream flow, even though the freestream is also irrotational. Generally, these two regions have curvature dependencies, which are further discussed in \S\ref{subsec:res_irrotational} and \S\ref{subsec:res_rotational}. The irrotational flow region was also observed above the turbulent wake for the uncontrolled flow not shown in figures~\ref{fig:Oz_mean} and \ref{fig:uv_mean}. The rigid-body rotation region, on the other hand, is specific to the controlled cases. As described in \S\ref{subsec:intro_curvature}, these two regions were also observed for reattached flows over a constant curvature body. Unlike the previous works, where the rigid-body rotation region appeared beneath the free surface and the turbulence levels in the rigid-body rotation region were only an order of magnitude higher than the irrotational flow region, the rigid-body rotation shear layer in the present study emerges near the wall and is highly turbulent, with the values of Reynolds shear stress in the shear layer being three orders of magnitude larger than the irrotational flow region. The similarities and differences of the two regions observed in the present work compared to the constant curvature geometries are further highlighted in \S\ref{subsec:res_irrotational} and \S\ref{subsec:res_rotational}.

\begin{figure}
  \centerline{\includegraphics[width=\linewidth, trim={2, 2, 2, 2}, clip]{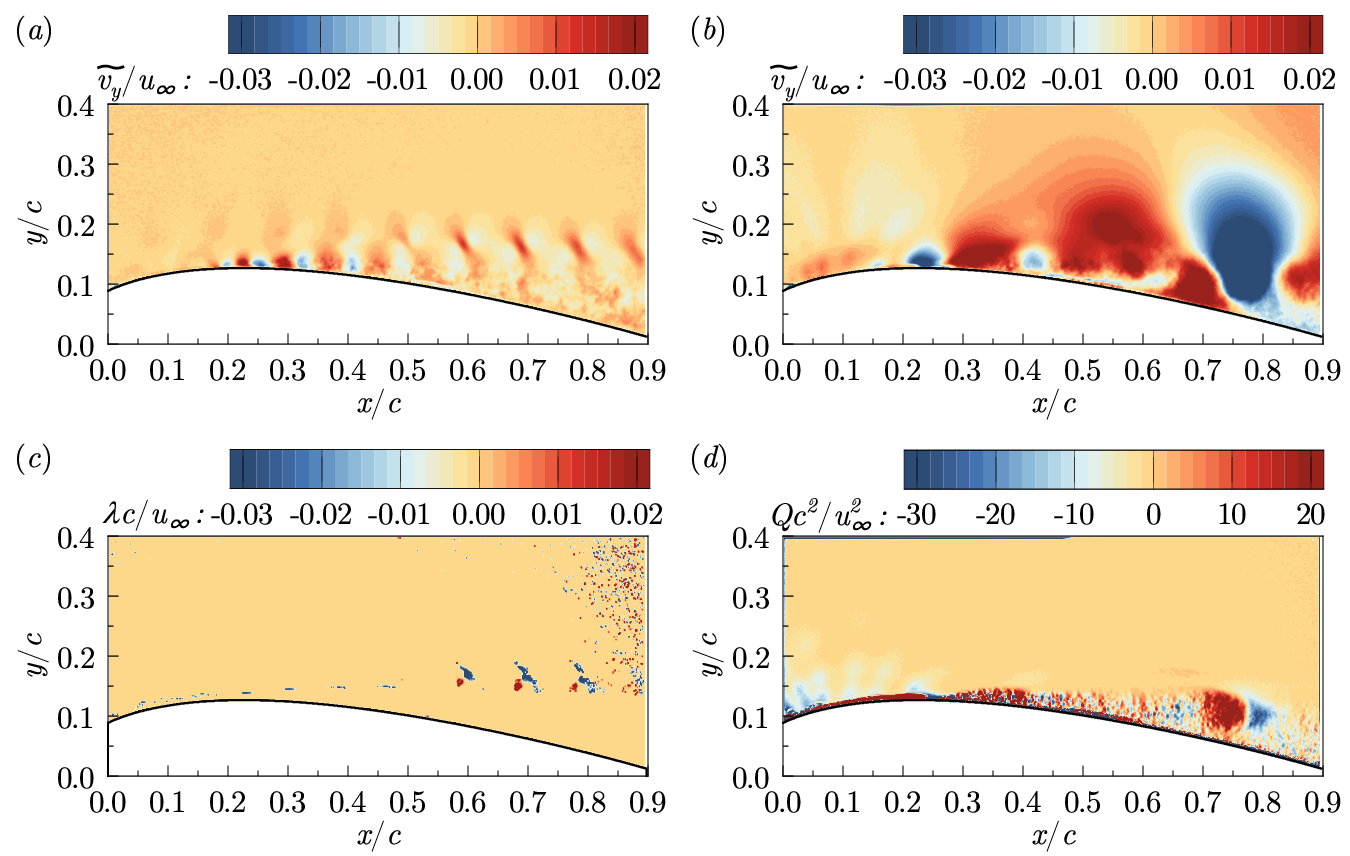}}
  \caption{Contour plots of (\textit{a}) and (\textit{b}) coherent vertical velocity for $F_c^+=11.53$ and $F_c^+=1.15$ cases, (\textit{c}) swirling strength for $F_c^+=11.53$ case, and (\textit{d}) $Q$-criterion for $F_c^+=1.15$ case, at the phase angle of $\phi=\ang{0}$.}
\label{fig:contours}
\end{figure}

From figures~\ref{fig:Oz_mean} and \ref{fig:uv_mean}, it is evident that the mean flow characteristics are very similar for $F_c^+=11.53$ and $F_c^+=1.15$ cases, confirming the observations of \citet{Xu2023}. Still, the previous works collectively indicate that the dynamics of the reattached flows should be noticeably different. Studying the dynamic aspects of the reattached flows over the airfoil model requires effective vortex identification methods. While certain vortex identifiers are only sensitive to small-scale structures, some are effective in detecting large-scale structures. A collection of vortex identification tools needed in the present study is provided in appendix~\ref{app:vortex_identification}. The goal of using these vortex identifiers is to understand the spatial distribution of the coherent structures with respect to the boundary layer, rigid-body rotation shear layer, and the irrotational flow region. The triple decomposition was applied to both controlled cases to visualize the large-scale turbulent structures. To capture the small-scale structures, the swirling strength and $Q$-criterion were applied to $F_c^+=11.53$ and $F_c^+=1.15$ controlled cases, respectively. Contour plots of the coherent component of the vertical velocity, swirling strength, and $Q$-criterion are shown in figure~\ref{fig:contours}. For $F_c^+=11.53$ case, both figures~\hyperref[fig:contours]{\ref{fig:contours}(a)} and \hyperref[fig:contours]{\ref{fig:contours}(c)} show a train of small-scale vortices that propagate away from the SJA array and grow in size as they evolve over the airfoil. The swirling strength identified these structures as tilted counter-rotating vortices with a slightly stronger downstream branch, characteristics that are very similar to the vortex rings described by \citet{Zhang2010} for circular SJAs operating in a crossflow. For $F_c^+=1.15$ case, the flow in the vicinity of the airfoil is strongly affected by the advection of large vortex structures as observed from figure~\hyperref[fig:contours]{\ref{fig:contours}(b)}. Still, $Q$-criterion detects some small-scale structures near the airfoil surface.

\begin{figure}
  \centerline{\includegraphics[width=\linewidth, trim={2, 2, 2, 2}, clip]{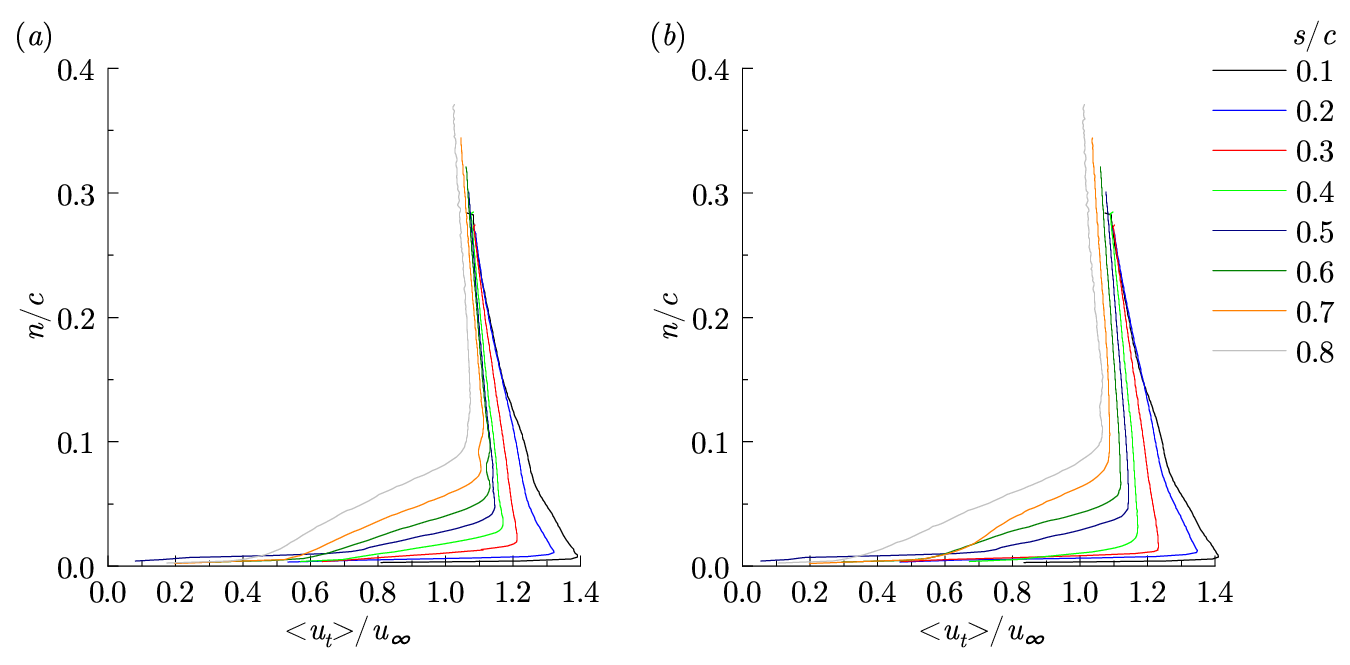}}
  \caption{Wall-normal profiles of the phase-averaged wall-tangent velocity at $\phi = \ang{90}$ along the $s$-constant lines for (\textit{a}) $F_c^+=11.53$ case and (\textit{b}) $F_c^+=1.15$ case.}
\label{fig:U_mean}
\end{figure}

The passage of vortex structures leaves a footprint on the phase-averaged velocity profiles. A sample of the $s$-constant profiles for the wall-tangent velocity averaged at $\phi = \ang{90}$ is presented in figure~\ref{fig:U_mean}. For $F_c^+=11.53$ case, the presence of the counter-rotating pairs shown in figure~\hyperref[fig:contours]{\ref{fig:contours}(c)} generates a local minimum and two peaks in the velocity profiles as may also be seen for the profiles at $s/c=0.6$, 0.7, and 0.8 in figure~\hyperref[fig:U_mean]{\ref{fig:U_mean}(a)}. Furthermore, comparing figures~\hyperref[fig:Oz_mean]{\ref{fig:Oz_mean}(a)} and \hyperref[fig:U_mean]{\ref{fig:U_mean}(a)}, it can be seen that these structures are advected along the boundary of the rigid-body rotation shear layer and the irrotational flow region, leaving the rigid-body rotation region mostly undisturbed. For $F_c^+=1.15$ case, the passage of large vortex structures generally leaves two different types of footprints on the velocity profiles. The first is a local minimum surrounded by two peaks occurring at the edge of the rigid-body rotation shear layer and the irrotational flow region, which can be seen for the profile at $s/c=0.8$ in figure~\hyperref[fig:U_mean]{\ref{fig:U_mean}(b)}. Compared to $F_c^+=11.53$ case, the distance between the two peaks is wider since the vortex structures are larger for $F_c^+=1.15$ case. The second is a sharp increase in the velocity occurring at the edge of the boundary layer and the rigid-body rotation shear layer as can be seen for $s/c=0.7$ in figure~\hyperref[fig:U_mean]{\ref{fig:U_mean}(b)}.

Another important flow property is the circulation $\Gamma$, related to the spanwise vorticity through the following equation:
\begin{equation}
\Gamma  = \iint{{\Omega _z}dS}
\label{eq:circulation}
\end{equation}
Using equation~\eqref{eq:circulation}, we may now confirm the observations of \citet{Amitay2002a}. For the rigid-body rotation shear layer and particularly at the edge of the boundary layer, upon reaching a steady-state condition, the circulation remains time-invariant for $F_c^+ \sim \mathcal{O}(10)$ case, whereas for $F_c^+ \sim \mathcal{O}(1)$ case it varies periodically. However, the mean spanwise vorticity, and consequently the mean circulation, is not significantly different between the two cases. Overall, for $F_c^+=1.15$ case, the passage of relatively larger vortical structures through the rigid-body rotation region disrupts the wall-normal balance of the vorticity. To evaluate the three-dimensionality of the mean flows at the centerline of the airfoil, a divergence test was performed by applying equation~\eqref{eq:divergence_test} to the Cartesian velocities:
\begin{equation}
\frac{{\partial \overline {{w_z}} }}{{\partial z}} =  - \frac{{\partial \overline {{u_x}} }}{{\partial x}} - \frac{{\partial \overline {{v_y}} }}{{\partial y}}
\label{eq:divergence_test}
\end{equation}
It was revealed that $\partial \overline {{w_z}}/{{\partial z}}$ is negligible in both the rigid-body rotation shear layer and the irrotational flow region, and is an order of $\mathcal{O}(10^{-1}{u_\infty }/c)$. Hence, the controlled flow in the two regions may be considered quasi-two-dimensional. A major objective of the present study is to characterize the flow characteristics in the two regions based on the variations of the curvature. To this end, we must solve the time-averaged spanwise vorticity and the continuity equation for an incompressible fluid in the airfoil curvilinear coordinate system, which follow immediately from equations~\eqref{eq:divergence} and \eqref{eq:curl}:
\begin{gather}
\frac{1}{{1 + \kappa n}}\frac{{\partial \overline {{u_t}} }}{{\partial s}} + \frac{1}{{1 + \kappa n}}\frac{{\partial (1 + \kappa n)\overline {{v_n}} }}{{\partial n}} + \frac{{\partial \overline {{w_z}} }}{{\partial z}} = 0 \label{eq:continuity_3c}\\
\overline {{\Omega _z}}  = \frac{1}{{1 + \kappa n}}\left( {\frac{{\partial \overline {{v_n}} }}{{\partial s}} - \frac{{\partial (1 + \kappa n)\overline {{u_t}} }}{{\partial n}}} \right) \label{eq:vorticity_3c}
\end{gather}
where $\kappa$ is the airfoil curvature. Based on the discussion so far, we have information about $\partial \overline {{w_z}}/{{\partial z}}$ and $\overline {{\Omega _z}}$ in the regions of interest, which are not enough to solve the system of equations. As shown in appendix~\ref{app:coordinate_transformation}, for a NACA 0025 airfoil, the normalized curvature gradient $c^2d\kappa/ds$ is an order of $\mathcal{O}(10)$ and $\mathcal{O}(1)$ within $s/c \lessapprox 0.3$ and $s/c \gtrapprox 0.3$, respectively. Still, the variations in the curvature may not be neglected anywhere, that is $d\kappa/ds \ne 0$. Moreover, assuming a constant curvature reduces the governing equations to those for a polar cylindrical coordinate system, entirely eliminating the curvature dependencies. Interestingly, by neglecting the term $\partial(1+\kappa n)/\partial s$ instead, which appears frequently in all governing equations in curvilinear coordinates, it is possible to maintain the curvature dependencies. To understand the rationale behind this assumption, consider the following dimensionless function:
\begin{equation}
\mathcal{K}(s,n) = c\frac{{\partial (1 + \kappa n)}}{{\partial s}} = \left( {\frac{n}{c}} \right)\left( {{c^2}\frac{{d\kappa }}{{ds}}} \right)
\label{eq:dimless_function}
\end{equation}
From equation~\eqref{eq:dimless_function}, it is clear that the term $\partial(1+\kappa n)/\partial s$ hinges on not only the curvature gradient $d\kappa/ds$ but also the wall-normal distance $n$. Notably, $\mathcal{K}$ is an order of magnitude smaller than $c^2d\kappa /ds$ within the irrotational flow region, where $n/c$ is an order of $\mathcal{O}(10^{-1})$. The order of magnitude difference becomes even greater in the rigid-body rotation shear layer since $n/c$ is an order of $\mathcal{O}(10^{-2})$. In general, there are two scenarios for which neglecting $\partial(1+\kappa n)/\partial s$ may cause large errors. The first scenario occurs for sufficiently large $n/c$ values. Indeed at very large distances from the wall, the flow is not affected by the airfoil and is predominantly in the $x$ direction. The second scenario happens when ${c^2}d\kappa /ds$ becomes very large, which is the case for $s/c \lessapprox 0.3$. Hence, so long as the multiplication of $n/c$ and ${c^2}d\kappa /ds$ is small enough, the assumption $\partial (1 + \kappa n)/\partial s \approx 0$ is sensible. In the subsequent analyses in \S\ref{subsec:res_irrotational} and \S\ref{subsec:res_rotational}, we shall neglect the term $\partial(1+\kappa n)/\partial s$ to simplify the system of governing equations.

\subsection{Irrotational flow region}\label{subsec:res_irrotational}
\begin{figure}
  \centerline{\includegraphics[width=\linewidth, trim={2, 2, 2, 2}, clip]{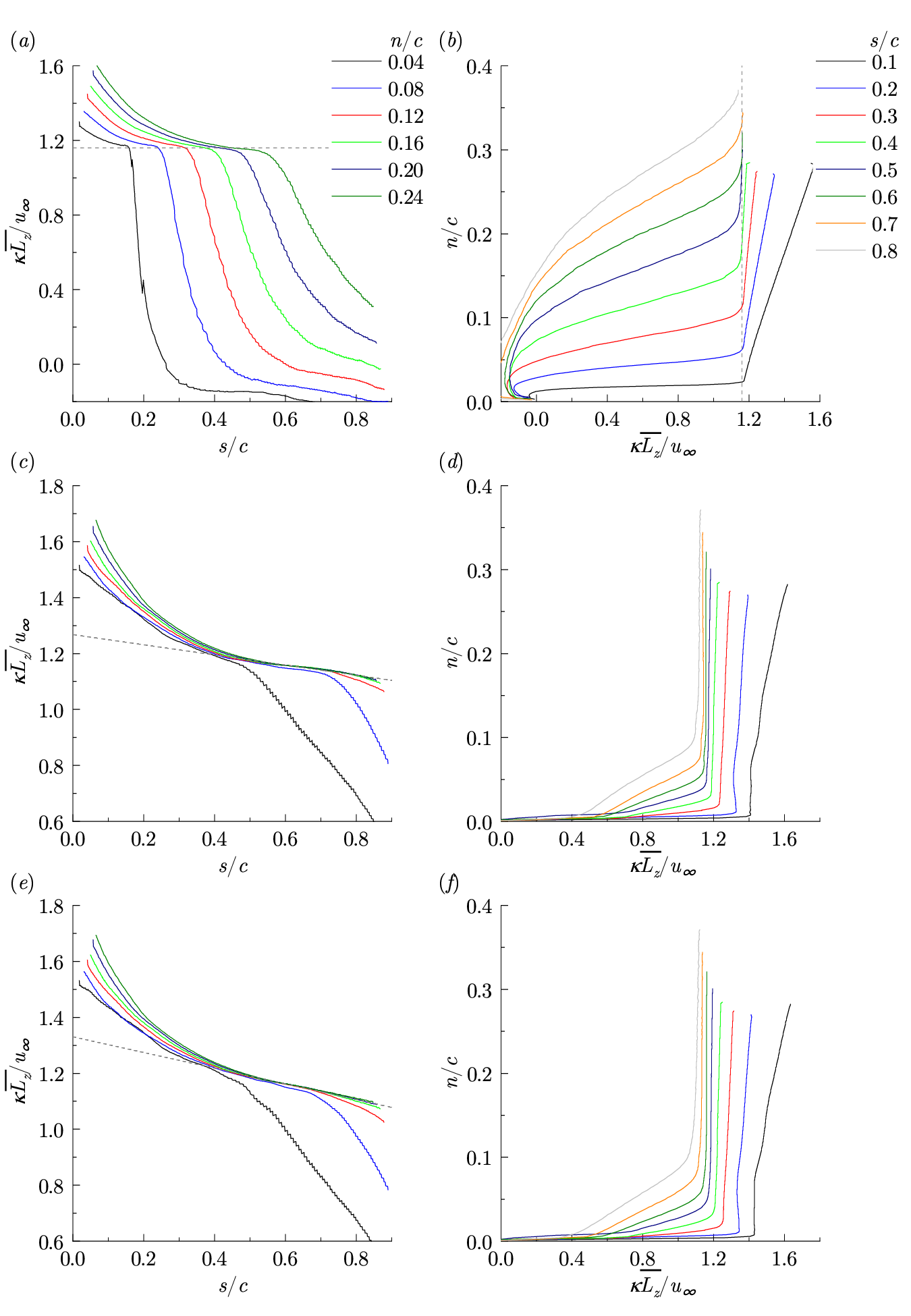}}
  \caption{Profiles of the curvature-multiplied mean angular momentum along the $n$- and $s$-constant lines. (\textit{a}), (\textit{c}), and (e) $n$-constant profiles, and (\textit{b}), (\textit{d}), and (f) $s$-constant profiles for the uncontrolled flow, $F_c^+=11.53$, and $F_c^+=1.15$ cases, respectively.}
\label{fig:Lz_mean}
\end{figure}

The angular momentum is among the most important rotational properties of curved fluid flows. In the airfoil curvilinear coordinate system, the angular momentum and the angular momentum multiplied by the curvature are defined as follows:
\begin{subequations}
\begin{align}
{L_z} &= (r_C+n){u_t} \label{eq:angular_moment}\\
\kappa {L_z} &= (1 + \kappa n){u_t} \label{eq:norm_angular_moment}
\end{align}
\end{subequations}
where $r_C$ is the radius of curvature. The profiles of the curvature-multiplied mean angular momentum along the $n$- and $s$-constant lines for all test cases are presented in figure~\ref{fig:Lz_mean}. It should be noted that the $n$- and $s$-constant lines may span multiple flow regions if not entirely situated within one region. Figures~\hyperref[fig:Lz_mean]{\ref{fig:Lz_mean}(a)}, \hyperref[fig:Lz_mean]{\ref{fig:Lz_mean}(c)}, and \hyperref[fig:Lz_mean]{\ref{fig:Lz_mean}(e)} indicate that the curvature-multiplied angular momentum gradually decreases in the wall-tangent direction within the irrotational region before exhibiting an abrupt decrease as the $n$-constant lines enter another region, that is the turbulent wake for the uncontrolled flow or the rigid-body rotation shear layer for the controlled cases. An interesting observation from figures~\hyperref[fig:Lz_mean]{\ref{fig:Lz_mean}(d)} and \hyperref[fig:Lz_mean]{\ref{fig:Lz_mean}(f)} is that for the controlled cases, the wall-normal profiles during the earlier stages of the flow development are comprised of two sections, an upper inclined section still not influenced by the curvature and a lower mostly upright section, corresponding to the freestream and irrotational flow regions, respectively. Once the flow evolves sufficiently in the tangential direction, the two sections merge into one fully developed profile that is almost invariant in the wall-normal direction. For the uncontrolled flow as shown in figure~\hyperref[fig:Lz_mean]{\ref{fig:Lz_mean}(b)}, negative values of the mean angular momentum replace the lower section of the wall-normal profiles due to the presence of the turbulent wake. Still, the upper inclined section of the profiles behaves similarly to that of the controlled cases. As discussed in \S\ref{subsec:intro_curvature}, the invariance of the mean angular momentum in the wall-normal direction was also observed for the irrotational flow region above the reattached flow on a constant curvature body and generally does not require axial symmetry. For clarity, the curvature-multiplied angular momentum profiles shown in figures~\hyperref[fig:Lz_mean]{\ref{fig:Lz_mean}(b)}, \hyperref[fig:Lz_mean]{\ref{fig:Lz_mean}(d)}, and \hyperref[fig:Lz_mean]{\ref{fig:Lz_mean}(f)} may be compared to the tangential velocity profiles presented in figure~\ref{fig:U_mean}. Evidently, multiplying the tangential velocity by a factor of $1+\kappa n$ has a straightening effect on the profiles, particularly for $s/c \gtrapprox 0.5$.

\begin{figure}
  \centerline{\includegraphics[width=\linewidth, trim={2, 2, 2, 2}, clip]{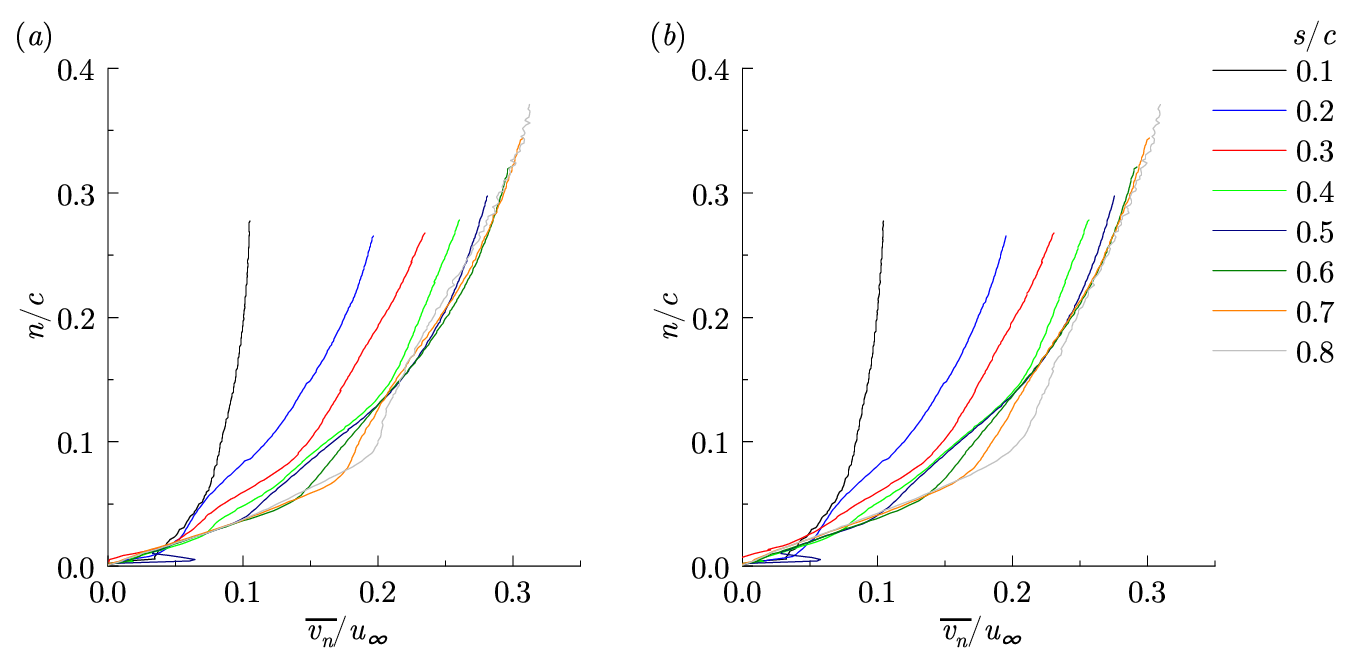}}
  \caption{Wall-normal profiles of the mean wall-normal velocity along the $s$-constant lines for (\textit{a}) $F_c^+=11.53$ case and (\textit{b}) $F_c^+=1.15$ case.}
\label{fig:V_mean}
\end{figure}

Based on the discussion in \S\ref{subsec:res_overview}, the flow may be considered quasi-two-dimensional within the irrotational flow region, that is $\partial \overline {{w_z}} /\partial z = 0$ and $\overline {{\Omega _z}}  = 0$. Now applying $\partial (1 + \kappa n)/\partial s \approx 0$, equations~\eqref{eq:continuity_3c} and \eqref{eq:vorticity_3c} may be rearranged as follows:
\begin{subequations}
\begin{align}
\frac{{\partial \overline {{w_z}} }}{{\partial z}} = 0 &\Rightarrow \frac{\partial }{{\partial s}}\left( {\frac{1}{{1 + \kappa n}}\frac{{\partial (1 + \kappa n)\overline {{u_t}} }}{{\partial s}}} \right) =  - \frac{{{\partial ^2}(1 + \kappa n)\overline {{v_n}} }}{{\partial s\partial n}} \label{eq:continuity_rearranged}\\
\overline {{\Omega _z}} = 0 &\Rightarrow \frac{\partial }{{\partial n}}\left( {(1 + \kappa n)\frac{{\partial (1 + \kappa n)\overline {{u_t}} }}{{\partial n}}} \right) = \frac{{{\partial ^2}(1 + \kappa n)\overline {{v_n}} }}{{\partial n\partial s}} \label{eq:vorticity_rearranged}
\end{align}
\end{subequations}
After applying Clairaut’s theorem (symmetry of the second partial derivatives) and summing equations~\eqref{eq:continuity_rearranged} and \eqref{eq:vorticity_rearranged}, we eliminate the wall-normal velocity to arrive at a governing equation for the mean angular momentum, which may be expressed using the Laplacian operator provided in equation~\eqref{eq:laplacian}:
\begin{subequations}
\begin{gather}
{\nabla ^2}\kappa \overline {{L_z}}  = 0 \label{eq:angular_laplace}\\
(1 + \kappa n)\frac{\partial }{{\partial n}}\left( {(1 + \kappa n)\frac{{\partial (1 + \kappa n)\overline {{u_t}} }}{{\partial n}}} \right) + \frac{{{\partial ^2}(1 + \kappa n)\overline {{u_t}} }}{{\partial {s^2}}} = 0 \label{eq:angular_momentum}
\end{gather}
\end{subequations}
When the curvature-multiplied mean angular momentum becomes invariant in the wall-normal direction, equation~\eqref{eq:angular_momentum} is reduced to:
\begin{subequations}
\begin{gather}
\kappa \overline {{L_z}}  = \mathcal{L}(s) \label{eq:angular_constant}\\
\frac{{{d^2}\mathcal{L}}}{{d{s^2}}} = 0 \Rightarrow \mathcal{L}(s) = A_{ir} s + B_{ir} \label{eq:momentum_linear}
\end{gather}
\end{subequations}
where $A_{ir}$ and $B_{ir}$ are unknown coefficients to be determined from experimental data. Here, the subscript $ir$ is used to denote the coefficients relating to the irrotational flow region. Equation~\eqref{eq:momentum_linear} describes a limiting behavior for the irrotational flow region, that is the curvature-multiplied mean angular momentum profiles become linear and collapse on the same line when the flow develops sufficiently over an airfoil. Even though the linear behavior of the curvature-multiplied angular momentum is apparent from figures~\hyperref[fig:Lz_mean]{\ref{fig:Lz_mean}(d)} and \hyperref[fig:Lz_mean]{\ref{fig:Lz_mean}(f)}, equation~\eqref{eq:momentum_linear} was fitted to the experimental data for $s/c \ge 0.5$ using all of the six n-constant lines shown in the figures to provide further evidence. A summary of the fitting parameters is provided in table~\ref{tab:fitting_angular}, with the fitted values depicted as gray dashed lines in figures~\hyperref[fig:Lz_mean]{\ref{fig:Lz_mean}(d)} and \hyperref[fig:Lz_mean]{\ref{fig:Lz_mean}(f)}. The high coefficients of determination $R^2$ after the fitting indicate a good agreement between the measured and predicted values. Finally, the distribution of the mean tangential velocity follows immediately by substituting equation~\eqref{eq:momentum_linear} in equation~\eqref{eq:norm_angular_moment}:
\begin{equation}
\overline {{u_t}} = \frac{{A_{ir} s + B_{ir}}}{{1 + \kappa n}} \label{eq:irrotational_u_velocity}
\end{equation}
By virtue of flow irrotationality and the invariance of the normalized angular momentum in the wall-normal direction, the mean wall-normal velocity should also become almost invariant in the wall-tangent direction near the trailing edge of the airfoil, that is $\partial \overline {{v_n}} /\partial s \approx 0$. This behavior may be seen from the $s$-constant profiles of the wall-normal velocity presented in figure~\ref{fig:V_mean}.

\newlength{\wt}
\setlength{\wt}{3cm}
\newcolumntype{z}[1]{>{\centering\arraybackslash\hspace{0pt}}m{#1}}
\begin{table}
\begin{center}
\begin{tabular}{z{\wt}z{\wt}z{\wt}z{\wt}}
Test case        & $A_{ir} c/u_\infty$       & $B_{ir} /u_\infty$      & $R^2$    \\
\parbox{1.65cm}{$F_c^+=11.53$} & -0.1815 & 1.2676 & 0.9662 \\
\parbox{1.65cm}{$F_c^+=1.15$}  & -0.2803 & 1.3303 & 0.9848
\end{tabular}
\caption{Summary of fitting parameters for equation~\eqref{eq:momentum_linear}.}
\label{tab:fitting_angular}
\end{center}
\end{table}

A linear variation in the mean angular momentum has also been observed above the reattached flow over a constant curvature body. The results of the present study clearly show that the linear behavior of the mean angular momentum for a constant curvature geometry may be generalized to the curvature-multiplied angular momentum for a varying curvature body under the circumstances explained earlier in \S\ref{subsec:res_overview}. The previous works also cover the analyses of the irrotational flow region over a recirculation bubble, where due to the finite reattachment length, the mean angular momentum above the recirculation bubble was characterized by a length scale. The presence of this characteristic length scale allows the Laplacian equation of the mean angular momentum to be solved through the separation of variables technique. Such a characteristic length does not exist for a turbulent wake since a wake continuously grows while entraining fluid from the above irrotational flow region. In contrast to a recirculation bubble, the present study shows that the curvature-multiplied mean angular momentum tends to remain constant above the turbulent wake, which is shown using a horizontal and a vertical gray dashed line in figures~\hyperref[fig:Lz_mean]{\ref{fig:Lz_mean}(a)} and \hyperref[fig:Lz_mean]{\ref{fig:Lz_mean}(b)}.

\subsection{Rigid-body rotation shear layer}\label{subsec:res_rotational}
\begin{figure}
  \centerline{\includegraphics[width=\linewidth, trim={2, 2, 2, 2}, clip]{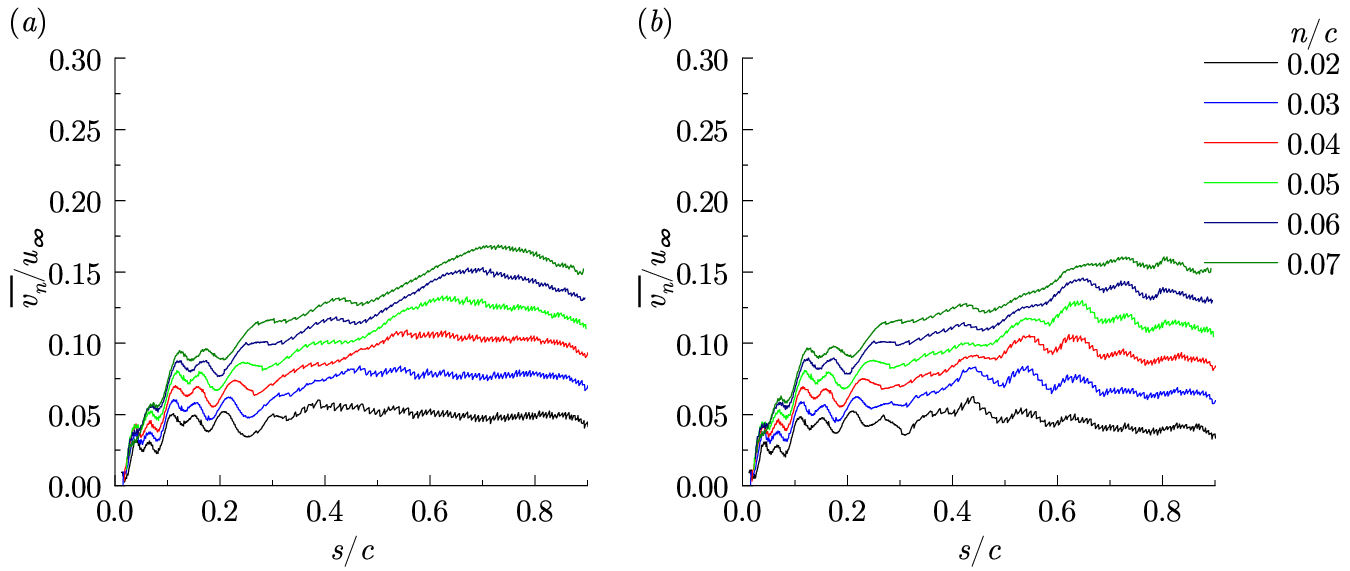}}
  \caption{Tangential profiles of the mean wall-normal velocity along the $n$-constant lines for (\textit{a}) $F_c^+=11.53$ case and (\textit{b}) $F_c^+=1.15$ case.}
\label{fig:V_mean_tangent}
\end{figure}

In the rigid-body rotation shear layer, which grows in thickness as the flow develops in the wall-tangent direction, the mean spanwise vorticity remains almost invariant in the wall-normal direction. In the present study, it was not possible to determine the exact onset of this shear layer due to the spatial resolution limitations. The distribution of the mean velocities in the rigid-body rotation region may be obtained from equations~\eqref{eq:continuity_3c} and \eqref{eq:vorticity_3c}. The invariance of the mean spanwise vorticity in the wall-normal direction, however, is not sufficient to solve the system of equations. For instance, even though the rigid-body rotation shear layer was also identified in the study of \citet{Shirinzad2023a}, no additional information was provided regarding the distribution of the mean spanwise vorticity or the mean velocities. Fortunately, the rigid-body rotation shear layer in the present work exhibits yet one more interesting feature that allows us to simplify equation~\eqref{eq:vorticity_3c}, that is $\partial \overline{v_n}/\partial s \approx 0$. Although this behavior may also be seen in figure~\ref{fig:V_mean}, it becomes much more apparent when the mean wall-normal velocity is observed along the $n$-constant lines as shown in figure~\ref{fig:V_mean_tangent}. From figure~\ref{fig:V_mean_tangent} it can be seen that the mean wall-normal velocity initially increases along the $n$-constant lines before reaching a stable value and remaining almost invariant for the section of the $n$-constant lines that lies within the rigid-body rotation region. By the invariance of the mean spanwise vorticity and the mean wall-normal velocity in the wall-normal and wall-tangent directions, equation~\eqref{eq:vorticity_3c} is reduced to:
\begin{subequations}
\begin{gather}
\frac{{\partial \overline {{v_n}} }}{{\partial s}} \approx 0 \label{eq:vn_constant}\\
\overline {{\Omega _z}}  =  - \frac{1}{{1 + \kappa n}}\frac{{d(1 + \kappa n)\overline {{u_t}} }}{{dn}} \label{eq:vorticity_reduced}
\end{gather}
\end{subequations}
Now integrating \eqref{eq:vorticity_reduced} gives rise to the distribution of the mean wall-tangent velocity within the rigid-body rotation region:
\begin{equation}
\overline {{u_t}}  =  - \frac{{1 + \kappa n}}{{2\kappa }}\overline {{\Omega _z}} (s) + \frac{{{\cal R}(s)}}{{1 + \kappa n}}
\label{eq:rotational_u_velocity}
\end{equation}
where the residual function $\mathcal{R}(s)$ emerges as a result of integration. It is now possible to fit mathematical expressions to the mean spanwise vorticity and the residual function to characterize the mean wall-tangent velocity.

Next, we shall try to find analytical forms for the mean spanwise vorticity and the residual function that are governed by the time-averaged continuity equation. To this end, we consider a quasi-two-dimensional flow and again apply $\partial (1 + \kappa n)/\partial s \approx 0$. Now substituting equation~\eqref{eq:rotational_u_velocity} in equation~\eqref{eq:continuity_3c}, we obtain:
\begin{equation}
\frac{{\partial \overline {{w_z}} }}{{\partial z}} = 0 \Rightarrow - \frac{{1 + \kappa n}}{2}\frac{d}{{ds}}\left( {\frac{{\overline {{\Omega _z}} }}{\kappa }} \right) + \frac{1}{{1 + \kappa n}}\frac{{d{\cal R}}}{{ds}} + \frac{\partial}{{\partial n}}(1 + \kappa n)\overline {{v_n}}  = 0
\label{eq:continuity_replaced}
\end{equation}
For the first term on the left-hand side of equation~\eqref{eq:continuity_replaced}, it is interesting to note that we were able to maintain the dependency of the mean spanwise vorticity on the curvature by neglecting $\partial (1 + \kappa n)/\partial s$ and not $d\kappa/ds$ as explained in \S\ref{subsec:res_overview}. Now for the left-hand-side of equation~\eqref{eq:continuity_replaced} to be zero, the first two terms can only be constant values. Hence, we can write out:
\begin{align}
\frac{d}{{ds}}\left( {\frac{{\overline {{\Omega _z}} }}{\kappa }} \right) = {A_{rr}} &\Rightarrow \overline {{\Omega _z}} (s) = \kappa ({A_{rr}}s + {B_{rr}}) \label{eq:rotational_vorticity}\\
\frac{{d\mathcal{R}}}{{ds}} = A_{re} &\Rightarrow \mathcal{R}(s) = {A_{re}}s + {B_{re}} \label{eq:rotational_int_function}
\end{align}
where $A_{rr}$, $A_{re}$, $B_{rr}$, and $B_{re}$ are empirical coefficients denoted using the $rr$ and $re$ subscripts to distinguish the rigid-body rotation and the residual parts, respectively. To evaluate these expressions, equations~\eqref{eq:rotational_vorticity} and \eqref{eq:rotational_int_function} were fitted to the experimental data for $s/c \ge 0.4$ using all the six n-constant lines shown in figure~\ref{fig:V_mean_tangent}. Experimental values for the residual function $\mathcal{R}(s)$ required for the data fitting were obtained by rearranging equation~\eqref{eq:rotational_u_velocity} as follows:
\begin{equation}
\mathcal{R}(s) = (1 + \kappa n)\overline {{u_t}}  + \frac{{{{(1 + \kappa n)}^2}}}{{2\kappa }}\overline {{\Omega _z}}
\end{equation}
A summary of the fitting parameters and the obtained coefficients of determination $R^2$ are shown in tables~\ref{tab:fitting_vorticity} and \ref{tab:fitting_function}. Equation~\eqref{eq:rotational_vorticity} also highlights the role of the wall curvature in the development of the rigid-body rotation shear layer. The formation of the rigid-body rotation shear layer for curved surfaces is clearly an artifact of the wall curvature since the mean spanwise vorticity becomes exactly zero for a flat plate with $\kappa=0$.

\begin{figure}
  \centerline{\includegraphics[width=\linewidth, trim={2, 2, 2, 2}, clip]{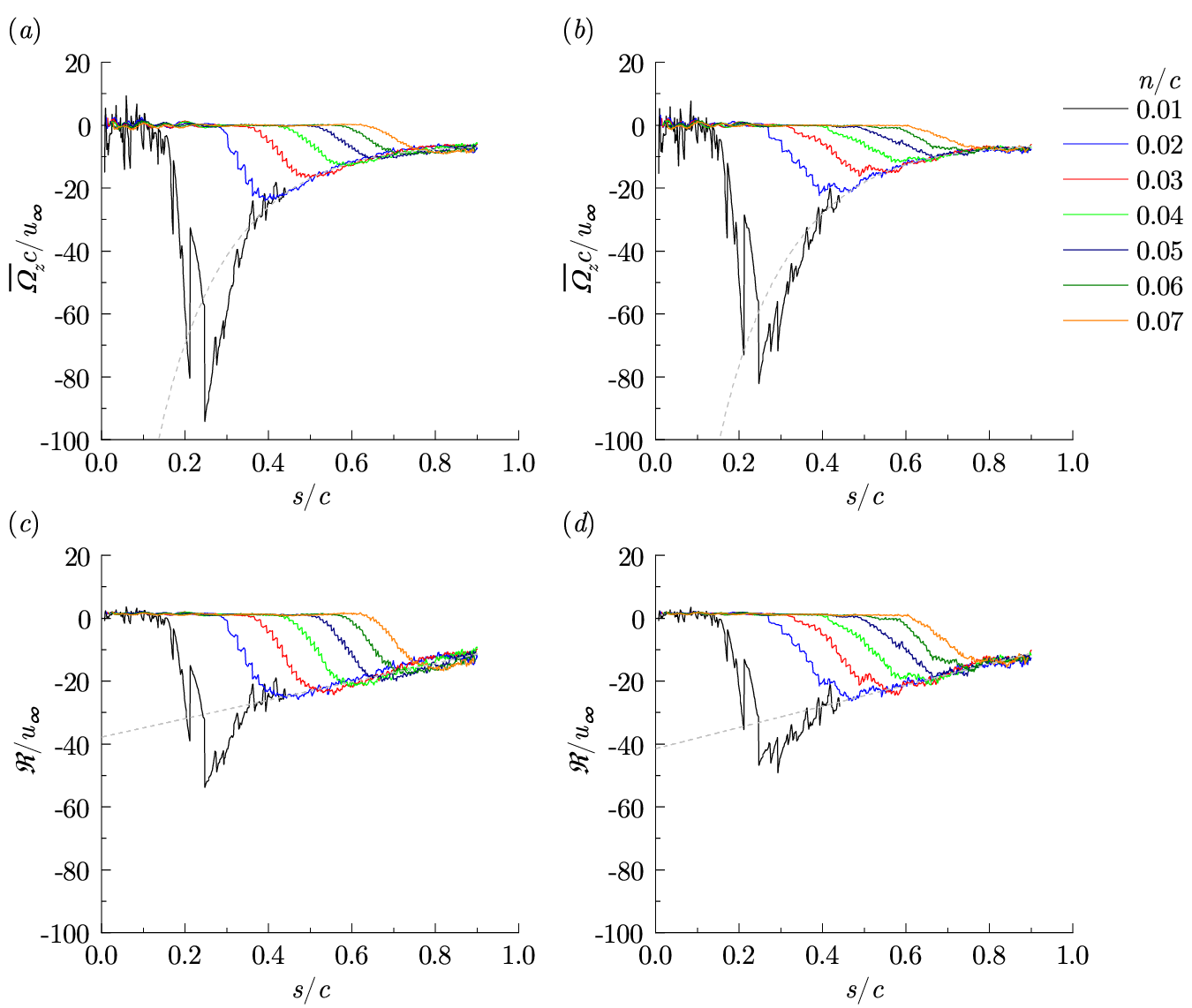}}
  \caption{Tangential profiles of the mean spanwise vorticity and the residual function along the $n$-constant lines. (\textit{a}) and (\textit{c}) $F_c^+=11.53$ case and (\textit{b}) and (\textit{d}) $F_c^+=1.15$ case. The fitted values to equations~\eqref{eq:rotational_vorticity} and \eqref{eq:rotational_int_function} are designated with a gray dashed line.}
\label{fig:Oz_mean_tangent}
\end{figure}

The $n$-constant profiles of the mean spanwise vorticity and the residual function within the rigid-body rotation region are presented in figure~\ref{fig:Oz_mean_tangent}, where the fitted values to equations~\eqref{eq:rotational_vorticity} and \eqref{eq:rotational_int_function} are depicted with gray dashed curves. Note that part of the $n/c=0.01$ profile lying within the boundary layer is not shown. A common feature of the profiles shown in figures~\hyperref[fig:Oz_mean_tangent]{\ref{fig:Oz_mean_tangent}(a)} and \hyperref[fig:Oz_mean_tangent]{\ref{fig:Oz_mean_tangent}(b)} is that the mean spanwise vorticity is initially zero for all $n$-constant lines. However, once the flow develops sufficiently in the wall-tangent direction, the irrotational flow is entrained in the shear layer, and after a small finite distance, the profiles follow a unified curve predicted by equation~\eqref{eq:rotational_vorticity}, continuously increasing in the wall-tangent direction. The $n$-constant profiles of the residual function $\mathcal{R}$ generally show similar behavior to the $n$-constant profiles of the mean spanwise vorticity. As may be seen from figures~\hyperref[fig:Oz_mean_tangent]{\ref{fig:Oz_mean_tangent}(c)} and \hyperref[fig:Oz_mean_tangent]{\ref{fig:Oz_mean_tangent}(d)}, the residual function has a zero value within the irrotational flow region and decreases suddenly as the flow enters the rigid-body rotation to follow the line described by equation~\eqref{eq:rotational_int_function}.

\begin{table}
\begin{center}
\begin{tabular}{z{\wt}z{\wt}z{\wt}z{\wt}}
Test case        & $A_{rr} c/u_\infty$       & $B_{rr}/u_\infty$      & $R^2$    \\
\parbox{1.65cm}{$F_c^+=11.53$} & 59.1962 & -76.7444 & 0.8845 \\
\parbox{1.65cm}{$F_c^+=1.15$}  & 69.2965 & -85.2202 & 0.9292
\end{tabular}
\caption{Summary of fitting parameters for equation~\eqref{eq:rotational_vorticity}.}
\label{tab:fitting_vorticity}
\end{center}
\rule{\textwidth}{0.5pt}
\vspace{2pt}
\begin{center}
\begin{tabular}{z{\wt}z{\wt}z{\wt}z{\wt}}
Test case        & $A_{re} c/u_\infty$       & $B_{re}/u_\infty$      & $R^2$    \\
\parbox{1.65cm}{$F_c^+=11.53$} & 29.3580 & -37.8277 & 0.8820 \\
\parbox{1.65cm}{$F_c^+=1.15$}  & 33.6912 & -41.5072 & 0.9219
\end{tabular}
\caption{Summary of fitting parameters for equation~\eqref{eq:rotational_int_function}.}
\label{tab:fitting_function}
\end{center}
\end{table}

\begin{figure}
  \centerline{\includegraphics[width=\linewidth, trim={2, 2, 2, 2}, clip]{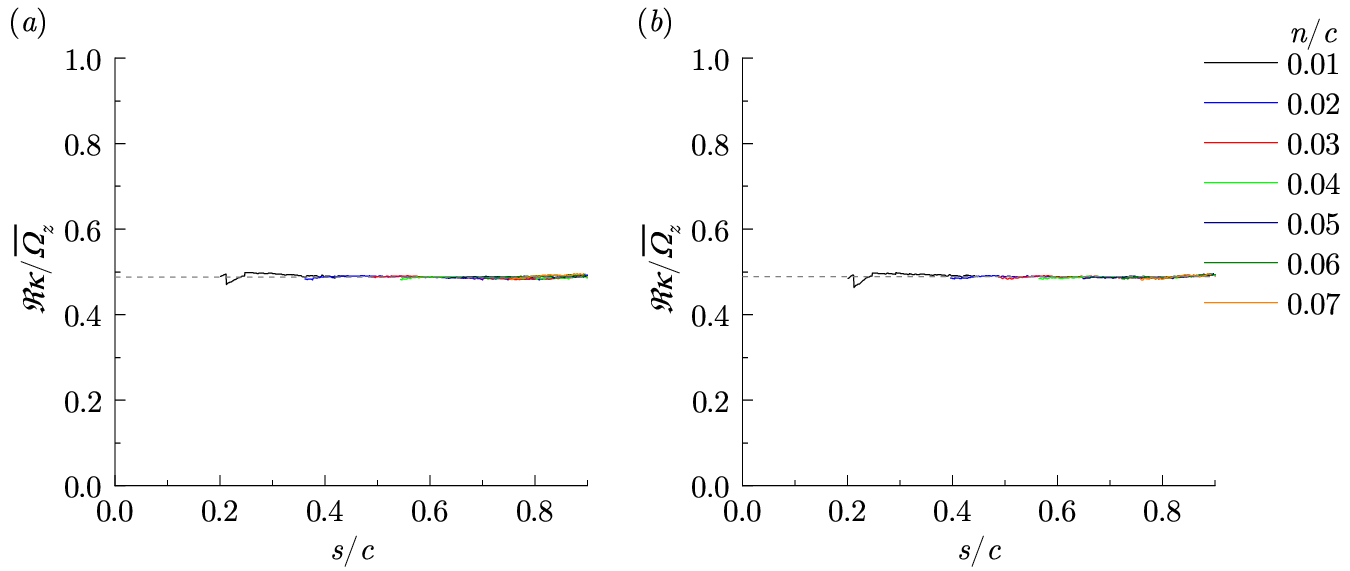}}
  \caption{Profiles of the residual fraction along the $n$-constant lines for (\textit{a}) $F_c^+=11.53$ case and (\textit{b}) $F_c^+=1.15$ case.}
\label{fig:F_tangent}
\end{figure}

A closer look at the coefficients presented in tables~\ref{tab:fitting_vorticity} and \ref{tab:fitting_function} reveals that the slope and intercept of the residual function are almost half of those for the mean spanwise vorticity, indicating that it may be possible to express the residual function as a fraction of the curvature-divided mean spanwise vorticity. The residual fraction therefore may be defined as follows:
\begin{equation}
\mathcal{F} = \frac{\mathcal{R} \kappa}{\overline{\Omega_z}}
\end{equation}
The profiles of the residual fraction $\mathcal{F}$ for the same $n$-constant lines shown in figure~\ref{fig:Oz_mean_tangent} are presented in figure~\ref{fig:F_tangent}. It should be noted that a section of the profiles situated within the irrotational flow region is emitted for better visualization. As expected, the residual fraction becomes almost a constant for both controlled cases within the rigid-body rotation shear layer, reaching a value of 0.4878 and 0.4885 for $F_c^+=11.53$ and $F_c^+=1.15$ cases, respectively. We may also represent the mean tangential velocity by incorporating the residual fraction as shown below:
\begin{equation}
\overline {{u_t}} = \left( {{A_{rr}}s + {B_{rr}}} \right)\left( { - \frac{{1 + \kappa n}}{2} + \frac{\mathcal{F}}{{1 + \kappa n}}} \right) \label{eq:rotational_uF_velocity}
\end{equation}

\begin{figure}
  \centerline{\includegraphics[width=\linewidth, trim={2, 2, 2, 2}, clip]{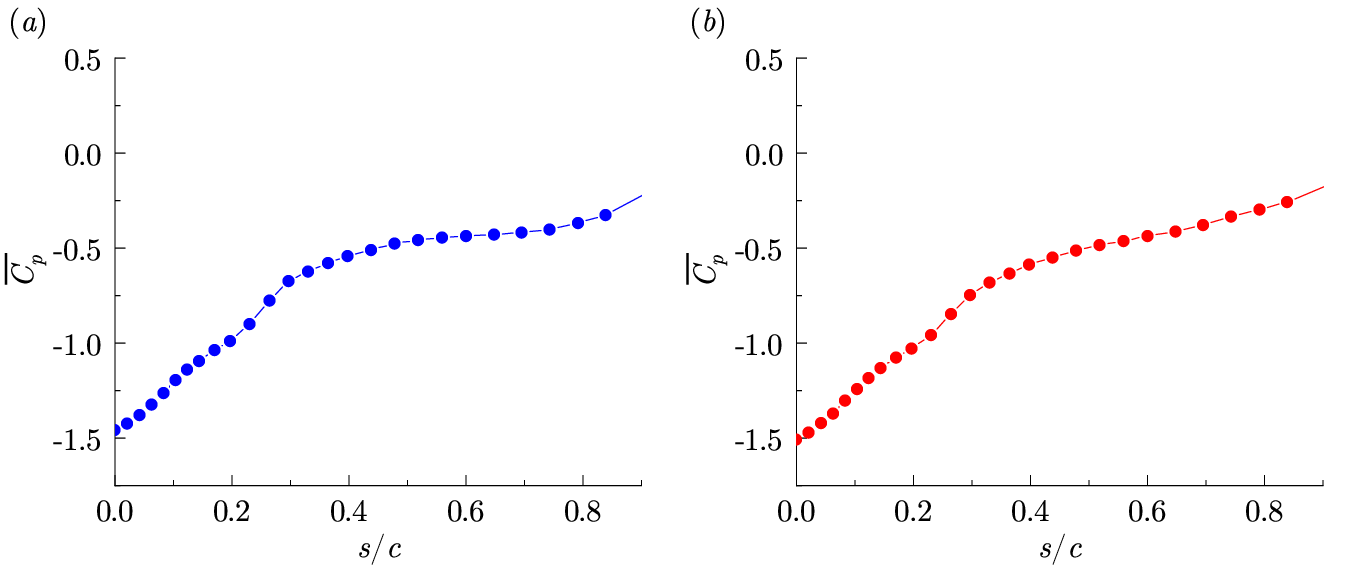}}
  \caption{One-dimensional profiles of the mean pressure coefficient along the airfoil surface for (\textit{a}) $F_c^+=11.53$ case and (\textit{b}) $F_c^+=1.15$ case.}
\label{fig:Cp_mean_tangent}
\end{figure}

For $s/c \gtrapprox 0.5$, the combined thickness of the rigid-body rotation region and the boundary layer as a function of the arc length may be obtained by equating equation~\ref{eq:irrotational_u_velocity} with equation~\ref{eq:rotational_uF_velocity}. Upon the formation of the rigid-body rotation shear layer, the reattached boundary layer no longer interacts with the irrotational flow region and instead has a common edge with the rigid-body rotation shear layer, where the edge tangential velocity is given by equations~\ref{eq:rotational_uF_velocity}, leading to a change in the behavior of the wall pressure. The mean pressure coefficient is given by:
\begin{equation}
\overline {{C_p}}  = \frac{{\bar p - {p_\infty }}}{{\dfrac{1}{2}\rho u_\infty ^2}}
\label{eq:pressure_coef}
\end{equation}
where $\rho$ is the air density and $p_\infty$ is the freestream pressure. The plots of the mean pressure coefficient for the controlled cases are presented in figure~\ref{fig:Cp_mean_tangent}. Evidently, the mean pressure coefficient continuously increases in the wall-tangent direction for both cases. As may also be seen from figures~\hyperref[fig:Cp_mean_tangent]{\ref{fig:Cp_mean_tangent}(a)} and \hyperref[fig:Cp_mean_tangent]{\ref{fig:Cp_mean_tangent}(b)}, the slope of the pressure coefficient is steeper for $s/c \lessapprox 0.3$ compared to $s/c \gtrapprox 0.3$, where the boundary layer interacts with the rigid-body rotation shear layer. As was highlighted in \S\ref{subsec:res_overview} and appendix~\ref{app:coordinate_transformation}, the values of the curvature gradient for $s/c \lessapprox 0.3$ and $s/c \gtrapprox 0.3$ are an order of magnitude apart, having a strong impact on the distribution of the wall pressure. Another interesting observation from figures~\hyperref[fig:Cp_mean_tangent]{\ref{fig:Cp_mean_tangent}(a)} and \hyperref[fig:Cp_mean_tangent]{\ref{fig:Cp_mean_tangent}(b)} is the sudden increase of the wall pressure near the trailing edge of the airfoil, indicating a deviation from rigid-body rotation dynamics as the flow enters the wake region.

\section{Summary and conclusions}\label{sec:summary}
The development of low-Reynolds number flows reattached using an array of circular microblowers over a NACA 0025 airfoil of chord length $c=\SI{300}{\milli\meter}$ at an angle of attack of $\alpha=\ang{10}$ and a chord-based Reynolds number of $Re_c=10^5$ was investigated experimentally for two distinct reduced frequencies $F_c^+=11.53$ and $F_c^+=1.15$, targeting the shear layer and wake instabilities. The measured velocity fields were decomposed along the axes of a curvilinear coordinate system that used the airfoil profile as its reference curve to specify the measurement points using the wall-normal distance $n$ and arc length $s$. Both $n$- and $s$-constant profiles of the time- and phase-averaged quantities, such as the mean spanwise vorticity, curvature-multiplied mean angular momentum, and Reynolds shear stress, were investigated to characterize the flow.

The wall-normal profiles of the curvature-multiplied mean angular momentum were examined for both the controlled and uncontrolled cases, where it was observed that the profiles become almost invariant in the wall-normal direction within the irrotational flow region after a sufficient development of the flow along the airfoil. Using the governing equations, it was shown that the invariance of the curvature-multiplied mean angular momentum in the wall-normal direction asserts that the $n$-constant profiles should vary linearly and collapse on the same line for the controlled cases. For the uncontrolled flow, the curvature-multiplied angular momentum was almost entirely constant within the irrotational flow region above the turbulent wake.

For both controlled cases, the wall-normal profiles of the mean spanwise vorticity and the Reynolds shear stress revealed the presence of a turbulent rigid-body rotation shear layer, stretching from the edge of the reattached boundary layer to the irrotational flow region and growing in thickness in the tangential direction. The $n$-constant profiles of the mean spanwise vorticity showed that the vorticity continuously increases in the wall-tangent direction within the rigid-body rotation region, faithfully following a unified curve. Using the governing equations, an analytical form was obtained for the variations of the mean tangential velocity and the spanwise vorticity, where it was shown that the curvature-divided mean spanwise vorticity varies linearly in the tangential direction. It was inferred that the rigid-body rotation shear layer is an artifact of the wall curvature, that is the mean spanwise vorticity becomes exactly zero for a flat plate with zero curvature.

Upon careful assessment of the phase-averaged profiles, it was revealed that the turbulent structures for $F_c^+=11.53$ case are advected along the boundary of the rigid-body rotation shear layer and the irrotational flow region. For the $F_c^+=1.15$ case, on the other hand, the passage of the large vortex structures through the rigid-body rotation region significantly disrupts the wall-normal balance of the vorticity. The distribution of the mean pressure coefficient along the airfoil surface highlighted the difference in the variations of the wall pressure for $s/c \lessapprox 0.3$ and $s/c \gtrapprox 0.3$. Compared to $s/c \lessapprox 0.3$, the pressure coefficient varied more gradually for $s/c \gtrapprox 0.3$, where the boundary layer was bounded by the rigid-body rotation shear layer. Near the trailing edge of the airfoil, a sudden increase in the wall pressure was detected, indicating a deviation from rigid-body rotation dynamics as the flow enters the wake region.

Since the flow was considered quasi-two-dimensional for both the irrotational flow region and the rigid-body rotation shear layer, a divergence test was conducted to show that $\partial \overline{w_z}/\partial z$ is negligible within the two flow regions. In addition, an order of magnitude analysis was conducted for $\mathcal{K} = c\partial{(1+\kappa n)}/\partial s$ to assess the validity of the assumption $\partial{(1+\kappa n)}/\partial s \approx 0$ needed to simplify the governing equations expressed in the curvilinear coordinate system for both the irrotational flow region and the rigid-body rotation shear layer. It was shown that $\mathcal{K}$ is at least an order of magnitude smaller than $c^2d\kappa/ds$, and consequently, it is reasonable to presume $\mathcal{K} \approx 0$ for $s/c \gtrapprox 0.3$, where $c^2d\kappa/ds$ was an order of $\mathcal{O}(1)$.

The authors acknowledge the support provided by the Natural Sciences and Engineering Research Council of Canada (NSERC), the Digital Research Alliance of Canada (the Alliance), and the University of Toronto.

\appendix
\section{Measurement uncertainty}\label{app:uncertainty}
This appendix is dedicated to the statistical convergence and quantification of the measurement uncertainty. The uncertainty in the PIV measurements depends on the bias and precision errors. In the present work, the bias errors, such as inaccurate estimation of the laser pulse time delay, optical distortion of the lenses, error in the camera magnification factor, and inaccurate response time of the tracer particles, were significantly reduced by carefully following the precautionary measures outlined by \citet{Prasad1992} and \citet{Forliti2000}. Symmetric translation of the interrogation windows was applied during the PIV image processing, which further contributes to reducing the bias error \citep{Meunier2003}. The precision errors, on the other hand, correspond to the computational errors in the statistical moments, which largely depend on the sample size $N$.

\begin{figure}
  \centerline{\includegraphics[width=\linewidth, trim={2, 2, 2, 2}, clip]{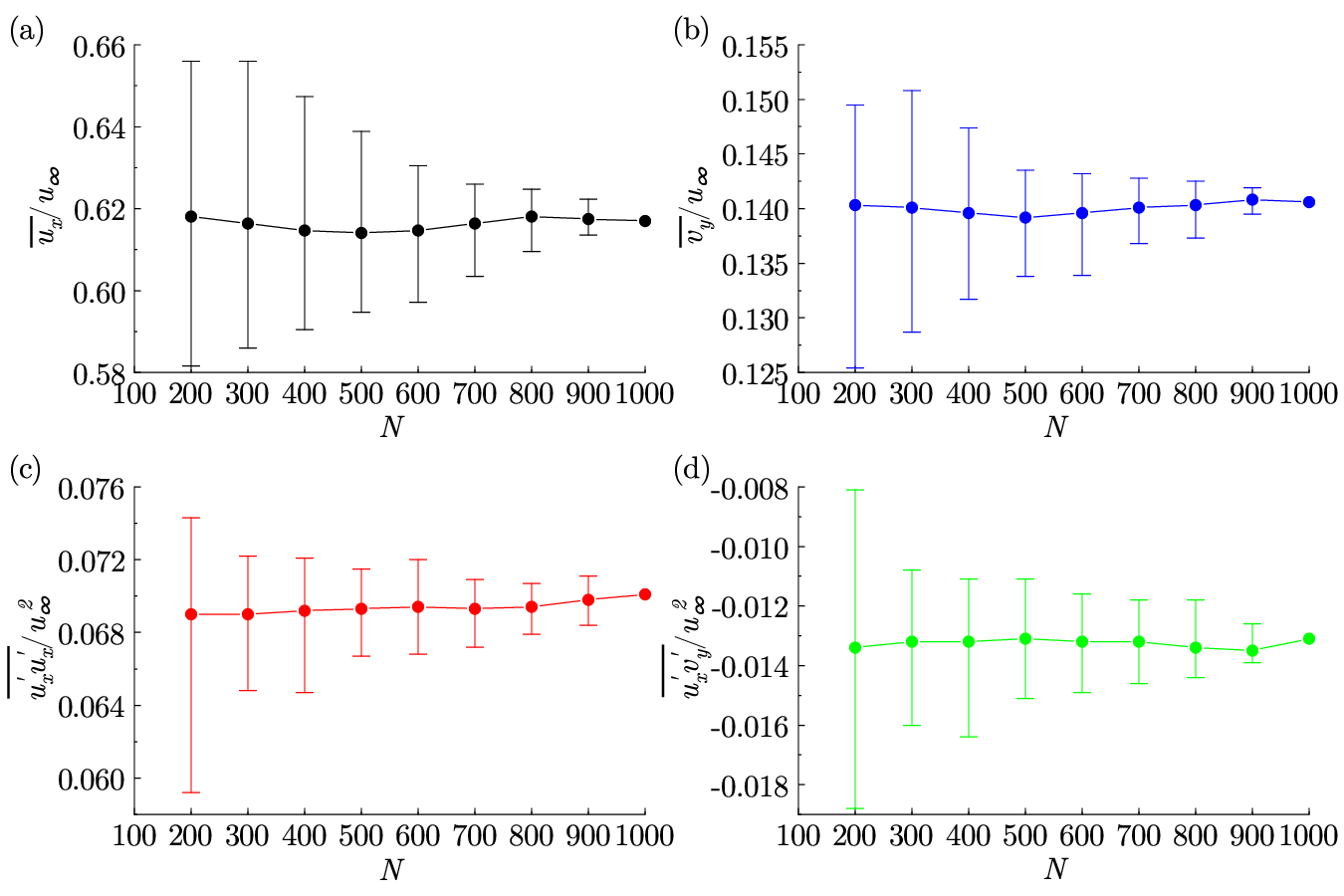}}
  \caption{Plots of the statistical convergence test at $x/c=0.4975$ and $y/c=0.2476$ for the uncontrolled flow.}
\label{fig:convergence_test}
\end{figure}

A statistical convergence test was performed using samples of different sizes, starting with $N=200$ and ending with $N=1000$ with an increment of 100, to calculate the first and second moments of statistics at a selected location within the separated shear layer for the uncontrolled flow. Except for $N=1000$, for which the statistics may be calculated only once, the process was repeated on all successive combinations of the instantaneous velocity fields to obtain the average values and the errors. For instance, for $N=900$ it was possible to calculate the statistics 100 times. The results of the convergence test with the error bars for the mean velocities and two of the Reynolds stresses are presented in figure~\ref{fig:convergence_test}. The samples were statistically independent since the PIV images were acquired in double-frame mode for the uncontrolled flow. From figure~\ref{fig:convergence_test} it can be seen that the relative error in the mean streamwise and transverse velocities for a sample size of $N=900$ are \SI{1.4}{\percent} and \SI{1.7}{\percent}, respectively. The relative errors in the Reynolds stresses, however, are an order of magnitude greater than that of the mean velocities. Hence, the mean velocities discussed in the present study are within a reasonable margin of error.

Following \citet{Sciacchitano} and \citet{Bendat2011}, the expressions for the measurement uncertainty are summarized below:
\begin{subequations}
\begin{align}
{\xi _{\overline {u_x} }} &= \frac{{{Z_C}}}{{\left| {\overline {u_x} } \right|}}\sqrt {\frac{{\overline {u_x'u_x'} }}{N}} \label{eq:uncertainty_u}\\
{\xi _{\overline {v_y} }} &= \frac{{{Z_C}}}{{\left| {\overline {v_y} } \right|}}\sqrt {\frac{{\overline {v_y'v_y'} }}{N}} \label{eq:uncertainty_v}\\
{\xi _{\overline {u_x'v_y'} }} &= {Z_C}\sqrt {\frac{{\overline {u_x'u_x'} \, \overline {v_y'v_y'}  + {{\overline {u_x'v_y'} }^2}}}{{(N - 1)\overline {u_x'u_x'} \, \overline {v_y'v_y'} }}} \label{eq:uncertainty_uv}
\end{align}  
\end{subequations}
where $Z_C$ represents the confidence coefficient, which may be set to $Z_C = 1.96$ to obtain the measurement uncertainties within \SI{95}{\percent} confidence level. The measurement uncertainties for the controlled cases at selected locations within the irrotational flow region and the rigid-body rotation shear layer obtained using $N=8000$ are reported in table~\ref{tab:uncertainty}. The uncertainty in the mean velocities and the Reynolds shear stress in the irrotational and rigid-body rotation regions is generally within $\pm \SI{4}{\percent}$.

\setlength{\wt}{2cm}
\begin{table}
\begin{center}
\begin{tabular}{z{\wt}z{\wt}z{\wt}z{\wt}z{\wt}z{\wt}}
\multicolumn{1}{c}{Test case} & $x/c$ & $y/c$ & $\xi_{\overline {u_x}}$ (\SI{}{\percent}) & $\xi_{\overline {v_y}}$ (\SI{}{\percent}) & $\xi_{\overline {u_x'v_y'}}$ (\SI{}{\percent}) \\
\multirow{2}{*}{\parbox{1.65cm}{$F_c^+=11.53$}}  & 0.8   & 0.08  & 0.37               & 3.22               & 2.39                  \\
                              & 0.8   & 0.20  & 0.06               & 0.79               & 2.20                  \\
\multirow{2}{*}{\parbox{1.65cm}{$F_c^+=1.15$}}   & 0.8   & 0.08  & 0.44               & 2.88               & 2.38                  \\
                              & 0.8   & 0.20  & 0.05               & 1.57               & 2.21                 
\end{tabular}
\caption{Measurement uncertainty quantification for the controlled cases at two selected locations within the irrotational flow region and the rigid-body rotation shear layer using $N = 8000$ samples.}
\label{tab:uncertainty}
\end{center}
\end{table}

\section{Curvilinear coordinates}\label{app:curvilinear_coordinates}
\begin{figure}
  \centerline{\includegraphics[width=\linewidth, trim={2, 2, 2, 2}, clip]{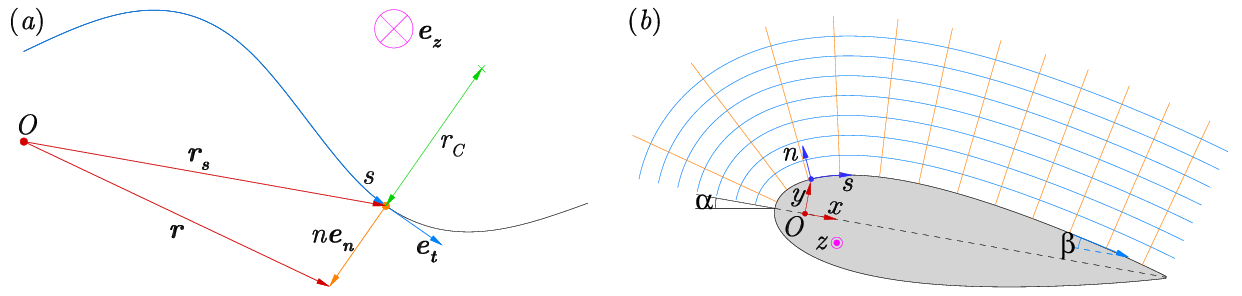}}
  \caption{Schematics showing (\textit{a}) Frenet-Serret frame for an arbitrary two-dimensional curve and (\textit{b}) adopted curvilinear coordinate system for an airfoil, with the $n$- and $s$-constant lines designated in blue and orange colors, respectively.}
\label{fig:nom_d}
\end{figure}

A curvilinear coordinate system is a coordinate system for Euclidean space in which the coordinate lines may be curved. Polar-cylindrical coordinates are an example of curvilinear coordinate systems, which have been frequently applied to study laminar and turbulent rotating flows \citep{Pope2000,Currie2016}. Consider an arbitrary two-dimensional curve $\boldsymbol{r_s}$ parameterized by its length $s$ as shown in figure~\hyperref[fig:nom_d]{\ref{fig:nom_d}(a)}, where $O$ is a fixed reference point in the plane. At a location $\boldsymbol{r_s}$ on the curve, the tangent, normal, and binormal unit vectors, denoted by $\boldsymbol{e_t}$, $\boldsymbol{e_n}$, and $\boldsymbol{e_z}$, are given by:
\begin{subequations}
\begin{align}
{\boldsymbol{e_t}} &= \frac{{d{\boldsymbol{r_s}}}}{{ds}} \label{eq:coordinates_frenet_a}\\
{\boldsymbol{e_n}} &= -\frac{1}{\kappa}\frac{{d{\boldsymbol{e_t}}}}{{ds}} \label{eq:coordinates_frenet_b}\\
{\boldsymbol{e_z}} &= {\boldsymbol{e_t}} \times {\boldsymbol{e_n}} \label{eq:coordinates_frenet_c}
\end{align}
\end{subequations}
where $\kappa$ is called the curvature, which is defined as the inverse of the radius of the circle tangent to the curve at the location $\boldsymbol{r_s}$ as shown in figure~\hyperref[fig:nom_d]{\ref{fig:nom_d}(a)}. Now from equations~\eqref{eq:coordinates_frenet_b} and \eqref{eq:coordinates_frenet_c} it follows that:
\begin{equation}
\frac{{d{\boldsymbol{e_n}}}}{{ds}} = \kappa {\boldsymbol{e_t}}
\label{eq:curvature}
\end{equation}
Equations~\eqref{eq:coordinates_frenet_b} and \eqref{eq:curvature} are collectively called the Frenet-Serret relations. In the curvilinear coordinates, the position of any point in the space may be expressed in terms of the normal distances $n$ and $z$ from the curve, that is:
\begin{equation}
\boldsymbol{r} = \boldsymbol{r_s} + n\boldsymbol{e_n} + z\boldsymbol{e_z}
\label{eq:position}
\end{equation}

\section{Coordinate transformation}\label{app:coordinate_transformation}
\begin{figure}
  \centerline{\includegraphics[width=\linewidth, trim={2, 2, 2, 2}, clip]{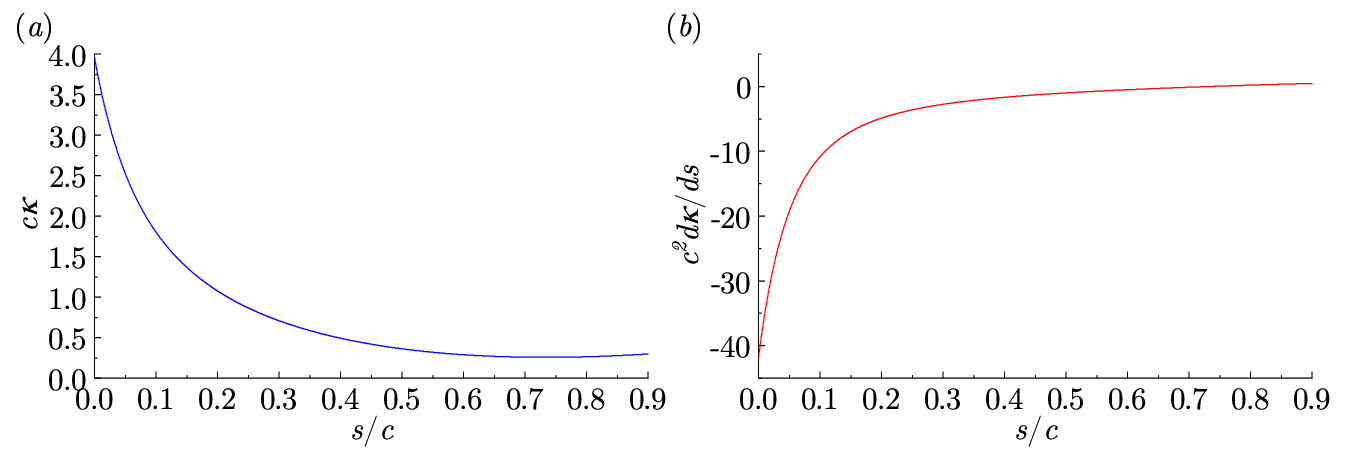}}
  \caption{Plots of (\textit{a}) curvature and (\textit{b}) gradient of curvature for a NACA 0025 airfoil.}
\label{fig:airfoil_curvature}
\end{figure}

For simplicity and performance reasons, the velocity measurements were performed in a Cartesian coordinate system with its origin $O$ at the left-bottom corner of the merged fields of view. The local coordinates are related to the global Cartesian coordinate system shown in figure~\ref{fig:nom_a} through equations~\eqref{eq:coordinates_local_a} and \eqref{eq:coordinates_local_b} as follows:
\begin{subequations}
\begin{align}
x &= X - X_O \label{eq:coordinates_local_a}\\
y &= Y - Y_O \label{eq:coordinates_local_b}
\end{align}
\end{subequations}
where $X_O/c=0.0741$ and $Y_O/c=0.0016$. The geometry of a NACA 0025 airfoil in the global Cartesian coordinate system is given by:
\begin{equation}
\frac{{\mathcal{C}(X)}}{c} = \frac{T}{{0.2}}\left( {{a_0}\sqrt {\frac{X}{c}}  + {a_1}\frac{X}{c} + {a_2}\frac{{{X^2}}}{{{c^2}}} + {a_3}\frac{{{X^3}}}{{{c^3}}} + {a_4}\frac{{{X^4}}}{{{c^4}}}} \right)
\label{eq:airfoil}
\end{equation}
The coefficients in equation~\eqref{eq:airfoil} are summarized in table~\ref{tab:airfoil_parameters}.  For an airfoil, a curvilinear coordinate system may be defined using the airfoil profile as the reference curve as shown in figure~\hyperref[fig:nom_d]{\ref{fig:nom_d}(b)}, where $s$ is the arc length, $n$ is the distance from the wall, and $z$ is the distance from the airfoil symmetry plane. The origin of the curvilinear coordinate system is at the left-most point on the surface of the airfoil within the fields of view, while the reference point $O$ is at the origin of the local Cartesian coordinate system.  The airfoil arc length is therefore given by:
\begin{equation}
s(x) = \bigintss_{X_O}^{X_O+x} {\sqrt {1 + {{\left( {\frac{{d\mathcal{C}}}{{dX}}} \right)}^2}} dX} 
\label{eq:arc_length}
\end{equation}
The curvature is obtained by substituting the parametrized airfoil profile given by equation~\eqref{eq:airfoil} in equation~\eqref{eq:coordinates_frenet_b}:
\begin{equation}
\kappa (x) = \frac{{\left| {\dfrac{{{d^2}\mathcal{C}}}{{d{x^2}}}} \right|}}{{{{\left( {1 + {{\left( {\dfrac{{d\mathcal{C}}}{{dx}}} \right)}^2}} \right)}^{\dfrac{3}{2}}}}}
\label{eq:airfoil_curvature}
\end{equation}
The plot of the curvature and its gradient with respect to the arc length is provided in figure~\ref{fig:airfoil_curvature}. An important observation from figure~\ref{fig:airfoil_curvature} is the difference in the order of magnitude of the curvature gradient for a NACA 0025 airfoil, that is $c^2 d \kappa/ds$ is an order of $\mathcal{O}(10)$ and $\mathcal{O}(1)$ for $s/c \lessapprox 0.3$ and $s/c \gtrapprox 0.3$, respectively.

\newcolumntype{Y}{>{\centering\arraybackslash}X}
\begin{table}
\begin{center}
\begin{tabularx}{\textwidth}{YYYYYY}
$T$  & $a_0$  & $a_1$   & $a_2$   & $a_3$  & $a_4$  \\
0.25 & 0.2969 & -0.1260 & -0.3516 & 0.2843 & -0.1015
\end{tabularx}
\caption{Open-trailing-edge NACA 0025 airfoil parameters.}
\label{tab:airfoil_parameters}
\end{center}
\end{table}

The velocities in the curvilinear coordinate system were obtained by transforming the Cartesian velocities through equations~\eqref{eq:transform_U} and \eqref{eq:transform_V}:
\begin{subequations}
\begin{align}
{u_t} = {u_x}\cos (\beta ) - {v_y}\sin (\beta ) \label{eq:transform_U}\\
{v_n} = {u_x}\sin (\beta ) + {v_y}\cos (\beta ) \label{eq:transform_V}
\end{align}
\end{subequations}
where $\beta$ is the slope of the airfoil surface with respect to the chord as shown in figure~\hyperref[fig:nom_d]{\ref{fig:nom_d}(b)}. Equations~\eqref{eq:transform_U} and \eqref{eq:transform_V} are also applicable to the mean and phase-averaged velocities. The transformation of the Reynolds shear stresses was achieved using equation~\eqref{eq:transform_uv} as follows:
\begin{equation}
\overline {u'_t v'_n} = \overline {u_x'v_y'} \cos (2\beta ) + \left( {\frac{{\overline {u_x'u_x'}  - \overline {v_y'v_y'} }}{2}} \right)\sin (2\beta ) \label{eq:transform_uv}
\end{equation}
For the pressure, the coordinates of the pressure taps are known in the global Cartesian coordinate system, which may easily be transformed by equation~\eqref{eq:arc_length} to obtain the surface pressure distribution along the airfoil surface.

\section{Vortex identification tools}\label{app:vortex_identification}
This appendix presents the structural analysis tools used in the current work.
For forced flows, since a phase angle may be defined based on a reference event in the flow, the Reynolds decomposition \citep{Reynolds1895} of the vertical velocity at a specific phase angle $\phi$ may be extended using the triple decomposition \citep{Hussain1970} according to:
\begin{subequations}
\begin{align}
{v_y} &= \overline {{v_y}} + v_y' = \overline {{v_y}}  + \left\langle {v_y'} \right\rangle + v_y'' \label{eq:triple_decompose}\\
\widetilde {\,{v_y}} &= \left\langle {v_y'} \right\rangle = \left\langle {v_y} \right\rangle - \overline {{v_y}} \label{eq:coherent_part}
\end{align}
\end{subequations}
where $\widetilde {\,{v_y}}$ and $v_y''$ are called the coherent and incoherent components, respectively. By definition, the spanwise vorticity $\Omega_z$, swirling strength $\lambda$, and $Q$-criterion are as follows \citep{Hunt1988,Zhou1999}:
\begin{gather}
\lambda = \left| {\im \left(
{\eig {\begin{bmatrix}
{\dfrac{{\partial {u_x}}}{{\partial x}}}&{\dfrac{{\partial {u_x}}}{{\partial y}}}\\
{\dfrac{{\partial {v_y}}}{{\partial x}}}&{\dfrac{{\partial {v_y}}}{{\partial y}}}
\end{bmatrix}} } \right)} \right| \times \sign({\Omega _z}) \label{eq:swirl_strength}\\
Q =  - \frac{1}{2}\left( {{{\left( {\frac{{\partial {u_x}}}{{\partial x}}} \right)}^2} + 2\left( {\frac{{\partial {u_x}}}{{\partial y}}} \right)\left( {\frac{{\partial {v_y}}}{{\partial x}}} \right) + {{\left( {\frac{{\partial {v_y}}}{{\partial y}}} \right)}^2}} \right) \label{eq:Q_criterion}
\end{gather}
Equations~ \eqref{eq:swirl_strength} and \eqref{eq:Q_criterion} apply equivalently to the time- and phase-averaged fields. From equation~\eqref{eq:swirl_strength} it is clear that the swirling strength has the advantage of detecting the direction of swirling motions compared to the $Q$-criterion.

\section{Governing equations}\label{app:governing_equations}
To express the governing equations in a curvilinear coordinate system, the gradient, Laplacian, divergence, and curl operators must be specified using the scale factors (also called Lame coefficients). For the airfoil geometry, these coefficients follow immediately from equations~\eqref{eq:coordinates_frenet_a}, \eqref{eq:curvature}, and \eqref{eq:position} as shown below \citep{Lewis1989}:
\begin{subequations}
\begin{align}
{h_s} &= \left| {\frac{{\partial {\boldsymbol{r}}}}{{\partial s}}} \right| = \left| {{{\boldsymbol{e}}_{\boldsymbol{t}}} + \kappa n{{\boldsymbol{e}}_{\boldsymbol{t}}}} \right| = 1 + \kappa n \label{eq:scale_s}\\
{h_n} &= \left| {\frac{{\partial \boldsymbol{r}}}{{\partial n}}} \right| = \left| {\boldsymbol{e_n}} \right| = 1 \label{eq:scale_n}\\
{h_z} &= \left| {\frac{{\partial \boldsymbol{r}}}{{\partial z}}} \right| = \left| {{\boldsymbol{e_z}}} \right| = 1 \label{eq:scale_z}
\end{align}
\label{eq:coordinates_frenet}
\end{subequations}
Once the scale factors are determined, the gradient, Laplacian, divergence, and curl operators in the curvilinear coordinate system are obtained by equations~\eqref{eq:gradient}, \eqref{eq:laplacian}, \eqref{eq:divergence}, and \eqref{eq:curl}:
\begin{align}
\bnabla &= \frac{1}{{{h_s}}}\frac{\partial }{{\partial s}}{\boldsymbol{e_t}} + \frac{1}{{{h_n}}}\frac{\partial }{{\partial n}}{\boldsymbol{e_n}} + \frac{1}{{{h_z}}}\frac{\partial }{{\partial z}}{\boldsymbol{e_z}} \label{eq:gradient}\\
{\nabla ^2} &= \frac{1}{{{h_s}{h_n}{h_z}}}\left( {\frac{\partial }{{\partial s}}\left( {\frac{{{h_n}{h_z}}}{{{h_s}}}\frac{{\partial}}{{\partial s}}} \right) + \frac{\partial }{{\partial n}}\left( {\frac{{{h_s}{h_z}}}{{{h_n}}}\frac{{\partial}}{{\partial n}}} \right) + \frac{\partial }{{\partial z}}\left( {\frac{{{h_s}{h_n}}}{{{h_z}}}\frac{{\partial}}{{\partial z}}} \right)} \right) \label{eq:laplacian}\\
\bnabla  \bcdot \boldsymbol{v} &= \frac{1}{{{h_s}{h_n}{h_z}}}\left( {\frac{\partial }{{\partial s}}({u_t}{h_n}{h_z}) + \frac{\partial }{{\partial n}}({v_n}{h_s}{h_z}) + \frac{\partial }{{\partial z}}({w_z}{h_s}{h_n})} \right) \label{eq:divergence} \\
\bnabla \times \boldsymbol{v} &= \frac{1}{{{h_s}{h_n}{h_z}}}\left| {\begin{matrix}
{{h_s}{\boldsymbol{e_t}}}&{{h_n}{\boldsymbol{e_n}}}&{{h_z}{\boldsymbol{e_z}}}\\
{\dfrac{\partial }{{\partial s}}}&{\dfrac{\partial }{{\partial n}}}&{\dfrac{\partial }{{\partial z}}}\\
{{u_t}{h_s}}&{{v_n}{h_n}}&{{w_z}{h_z}}
\end{matrix}} \right| \label{eq:curl}
\end{align}
Particularly, the continuity equation for an incompressible flow is simply $\bnabla \bcdot \boldsymbol{v}=0$. The vorticity is also defined as the curl of the velocity and is denoted by $\boldsymbol{\Omega} = \bnabla \times \boldsymbol{v}$.

\bibliographystyle{jfm}
\phantomsection
\addcontentsline{toc}{section}{References}
\bibliography{references}

\end{document}